\documentclass{JHEP}
\usepackage{epsfig}
\newcommand{\eref}[1]{(\ref{#1})}

\newcommand{\fref}[1]{Figure~\ref{#1}}
\newcommand{\cref}[1]{Chapter~\ref{#1}}
\newcommand{\beq}{\begin{equation}}
\newcommand{\eeq}{\end{equation}}
\newcommand{\ba}{\begin{array}}
\newcommand{\ea}{\end{array}}
\newcommand{\bcenter}{\begin{center}}
\newcommand{\ecenter}{\end{center}}

\def\IB{\relax\hbox{$\inbar\kern-.3em{\rm B}$}}
\def\IC{\relax\hbox{$\inbar\kern-.3em{\rm C}$}}
\def\ID{\relax\hbox{$\inbar\kern-.3em{\rm D}$}}
\def\IE{\relax\hbox{$\inbar\kern-.3em{\rm E}$}}
\def\IF{\relax\hbox{$\inbar\kern-.3em{\rm F}$}}
\def\IG{\relax\hbox{$\inbar\kern-.3em{\rm G}$}}
\def\IGa{\relax\hbox{${\rm I}\kern-.18em\Gamma$}}
\def\IH{\relax{\rm I\kern-.18em H}}
\def\IK{\relax{\rm I\kern-.18em K}}
\def\IL{\relax{\rm I\kern-.18em L}}
\def\IP{\relax{\rm I\kern-.18em P}}
\def\IR{\relax{\rm I\kern-.18em R}}
\def\IZ{\relax\ifmmode\mathchoice
{\hbox{\cmss Z\kern-.4em Z}}{\hbox{\cmss Z\kern-.4em Z}}
{\lower.9pt\hbox{\cmsss Z\kern-.4em Z}}
{\lower1.2pt\hbox{\cmsss Z\kern-.4em Z}}\else{\cmss Z\kern-.4em Z}\fi}
\def\II{\relax{\rm I\kern-.18em I}}

\def\sCC{{\kern 0.27em\vrule height1.45ex width0.03em depth0em
          \kern-0.30em\rm C}}
\def\C{{\mathchoice
  {\sCC}
  {\sCC}
  {\kern 0.225em \vrule height1.05ex width0.025em depth0em \kern-0.25em \rm C}
  {\kern 0.180em \vrule height0.78ex width0.02em depth0em \kern-0.2em \rm C}
        }}
\def\sHH{{\rm I\kern-.16em{}H}}
\def\H{{\mathchoice
  {\sHH}
  {\sHH}
  {\rm I\kern-.13em{}H}
  {\rm I\kern-.13em{}H} }}
\def\sNN{{\rm I\kern-.16em{}N}}
\def\N{{\mathchoice
  {\sNN}
  {\sNN}
  {\rm I\kern-.12em{}N}
  {\rm I\kern-.10em{}N} }}
\def\sPP{{\rm I\kern-.16em{}P}}
\def\P{{\mathchoice
  {\sPP}
  {\sPP}
  {\rm I\kern-.12em{}P}
  {\rm I\kern-.10em{}P} }}
\def\sQQ{{\kern 0.27em \vrule height1.45ex width0.03em depth0em
          \kern-0.30em \rm Q}}
\def\Q{{\mathchoice
        {\sQQ}
        {\sQQ}
  {\kern 0.225em \vrule height1.05ex width0.025em depth0em \kern-0.25em \rm Q}
  {\kern 0.180em \vrule height0.78ex width0.020em depth0em \kern-0.20em \rm Q}
        }}
\def\sRR{{\rm I\kern-0.16em{}R}}
\def\R{{\mathchoice
  {\sRR}
  {\sRR}
  {\rm I\kern-0.12em{}R}
  {\rm I\kern-0.10em{}R} }}
\def\sZZ{{\rm Z\kern-0.32em{}Z}}
\def\Z{{\mathchoice
  {\sZZ}
  {\sZZ} 
  {\rm Z\kern-0.3em{}Z}     %.3
  {\rm Z\kern-0.25em{}Z} }}  %.25
\def\ZZZ{{\rm Z\kern-0.24em{}Z}}
\def\sII{{\rm I\kern-0.16em{}I}}
\def\I{{\mathchoice
  {\sII}
  {\sII}
  {\rm I\kern-0.12em{}I}
  {\rm I\kern-0.10em{}I} }}

\def\inbar{\,\vrule height1.5ex width.4pt depth0pt}
\font\cmss=cmss10 \font\cmsss=cmss10 at 7pt

\def\odd{{\rm odd}}
\def\even{{\rm even}}

\def\smiley{\hbox{\large$\bigcirc$\hspace{-0.80em}\raise.2ex
\hbox{$\cdot\cdot$}\kern-.61em\lower.2ex\hbox{\scriptsize$\smile$}}\ }
\def\frowny{\hbox{\large$\bigcirc$\hspace{-0.80em}\raise.2ex
\hbox{$\cdot\cdot$}\kern-.635em\lower.2ex\hbox{\scriptsize$\frown$}}\ }

\def\I{{\rlap{1} \hskip 1.6pt \hbox{1}}}

\newcommand{\gen}[1]{\langle #1 \rangle}

\makeatletter
\let\hangafter\@hangfrom
\makeatother

\newtheorem{definition}{\sf DEFINITION}

\newtheorem{theorem}{\sf THEOREM}

\preprint{MIT-CTP-3046\\ \\ {\tt hep-th/}}

\title{Discrete Torsion, Covering Groups \\and Quiver Diagrams}
\author{Bo Feng, Amihay Hanany, Yang-Hui He and Nikolaos Prezas
\footnote{
Research supported in part
by the Reed Fund Award, 
the CTP and the LNS of MIT and the U.S. Department of Energy 
under cooperative research agreement \# DE-FC02-94ER40818.
A. H. is also supported by an A. P. Sloan Foundation Fellowship,
and a DOE OJI award.}
\\
Center for Theoretical Physics,
\\ Massachusetts Institute of Technology,\\ Cambridge, MA 02139, USA.\\
\email{fengb, hanany, yhe, prezas@ctp.mit.edu}
}
\abstract{Without recourse to the sophisticated machinery
of twisted group algebras,
projective character tables and explicit values of 2-cocycles, we here
present a simple algorithm to study the gauge theory data of D-brane probes
on a generic orbifold $G$ with discrete torsion turned on.
We show in particular
that the gauge theory can be obtained with the knowledge of no more
than the {\em ordinary} character tables of $G$ and its covering group
$G^*$. Subsequently we present the quiver diagrams of certain
illustrative examples of
$SU(3)$-orbifolds which have non-trivial Schur
Multipliers. The paper serves as a companion to our
\href{http://xxx.lanl.gov/abs/hep-th/0010023}{earlier
work} and aims to initiate a systematic and computationally
convenient study of discrete torsion.}
\begin{document}
\newpage
\section{Introduction}
Discrete torsion \cite{Vafa,VafaWit} has become a meeting ground for
many interesting
sub-fields of string theory; its intimate relation with background
B-fields and non-commutative geometry is one of its many salient
features. In the context of D-brane probes on orbifolds with discrete
torsion turned on, new classes of gauge theories may be fabricated and
their (non-commutative) moduli spaces, investigated (see from
\cite{Doug} to \cite{Bantay}). Indeed, as it was
pointed out in \cite{Doug}, projection on the matter spectrum in the
gauge theory by an orbifold $G$ with non-trivial discrete torsion
is performed by the {\em projective representations} of $G$, 
rather than the mere linear (ordinary) represenations as in the case
without.

In a previous paper \cite{FHHP}, to which the present work shall be a
companion, we offered a classification of the orbifolds with ${\cal
N}=0,1,2$ supersymmetry which permit the turning on of discrete torsion.
We have pointed there that for the orbifold group $G$, the discriminant
agent is the Abelian group known as the {\bf Schur Multiplier} $M(G) :=
H^2(G,\IC^*)$; only if $M(G)$ were non-trivial could $G$ afford a
projective representation and thereby discrete torsion.

In fact one can do more and for actual physical computations one needs
to do more. The standard procedure of calculating the matter content
and superpotential of the orbifold gauge theory as developed in
\cite{LNV} can, as demonstrated in \cite{AspinPles}, be directly
generalised to the case with discrete torsion. Formulae given in terms
of the ordinary characters have their immediate counterparts in terms of
the projective characters, the {\it point d'appui} being that the
crucial properties of ordinary characters, notably orthogonality,
carry over without modification, to the projective case.

And thus our task would be done if we had a method of computing the
projective characters. Upon first glance, this perhaps seems
formidable: one seemingly is required to know the values of the cocycle
representatives $\alpha(x,y)$ in $M(G)$ for all $x,y \in G$. In
actuality, one can dispense with such a need. There exists a canonical
method to arrive at the projective characters, namely by recourse to
the {\bf covering group} of $G$. We shall show in this writing
the methodology standard in the mathematics
literature \cite{Karp,Humph} by which one, once
armed with the Schur Multiplier, arrives
at the cover. Moreover, in light of the physics, we will show how,
equipped with no more than the knowledge of the character table of $G$
and that of its cover $G^*$, one obtains the matter content of
the orbifold theory with discrete torsion.

The paper is organised as follows. Section 2 introduces the
necessary mathematical background for our work. Due to the
technicality of the details, we present a paragraph at the beginning 
of the section to summarise the useful facts; the reader may then
freely skip the rest of Section 2 without any loss.
In Section 3, we commence with an explicit example, viz., the
ordinary dihedral group, to demonstrate the method to construct the
covering group. Then we present all the covering groups for transitive
and intransitive discrete subgroups of $SU(3)$.
In Section 4, we use these covering groups
to calculate the corresponding gauge 
theories (i.e., the quiver diagrams) for all exceptional subgroups
of $SU(3)$ admitting discrete torsion as well as some examples for the
Delta series. In
particular we demonstrate the algorithm of extracting the quivers from
the ordinary character tables of the group and its cover.
As a by-product, in Section 5 we present a method to calculate the
{\em cocycles} directly which will be useful for future reference.
The advantage of our methods for the quivers and the cocycles is
their simplicity and generality. Finally, in Section 6 we give some
conclusions and further directions for research.
\section*{Nomenclature}
Throughout this paper, unless otherwise specified, we shall adhere to
the following conventions for notation:

\begin{tabular}{rl}
$\omega_n$ & $n$-th root of unity;\\
$G$ & a finite group of order $|G|$;\\
$[x,y]$ &  $:= xyx^{-1}y^{-1}$, the group commutator of $x,y$;\\
$\gen{x_i|y_j}$ & the group generated by elements $\{x_i\}$ with
        relations $y_j$;\\ 
$\gcd(m,n)$ & the greatest common divisor of $m$ and $n$; \\
$Z(G)$ & centre of  $G$;\\
$G':=[G,G]$ & the derived (commutator) group of $G$;\\
$G^*$ & the covering group of $G$;\\
$A=M(G)$ & the Schur Muliplier of $G$;\\
char$(G)$ & the ordinary (linear) character table of $G$, given as an
$(r + 1) \times r$ matrix \\
 & with $r$ the number of conjugacy classes and
	the extra row for the class numbers;\\
$Q_\alpha(G,{\cal R})$ & the $\alpha$-projective quiver for $G$ associated to
	the chosen representation ${\cal R}$.
\end{tabular}
\section{Mathematical Preliminaries}
We first remind the reader of some properties of the the theory of
projective representations; in what follows we adhere to the notation
used in our previous work \cite{FHHP}.

Due to the technicalities in the ensuing, the audience might be distracted
upon the first reading. Thus as promised in the introduction, we here
summarise the keypoints in the next fews paragraphs, so that the
remainder of this section may be loosely perused without any loss.

Our aim of this work is to attempt to construct the gauge theory
living on a D-brane probing an orbifold $G$ when
``discrete torsion'' is turned on. To accomplish such a goal, we need
to know the projective representations of the finite group $G$, which
may not be immediately available.
However, mathematicians have shown that there exists (for
representations in $GL(\IC)$) a
group $G^*$ called the {\em covering group} of $G$, such that
there is a one-to-one correspondence between the projective
representations of $G$ and the linear (ordinary) representations of
$G^*$.
Thus the method is clear: we simply need to find the covering group and
then calculate the ordinary characters of its (linear) representations.

More specificaly, we first introduce the concept of the
covering group in Definition 2.2. Then in Theorem 2.1, we introduce
the necessary and sufficient conditions for $G^*$ to be a covering
group; these conditions are very important and we use them
extensively during actual computations.

However, $G^*$ for any given $G$ is not unique and there
exist non-isomorphic groups which all serve as covering groups.
To deal with this, we introduce {\em isoclinism} and show that these
non-isomorphic covering groups must be isoclinic to each other in
Theorem 2.2. Subsequently, in Theorem 2.3, we give an upper-limit on 
the number of non-isomorphic covering groups of $G$. 
Finally in Thereom 2.4 we present
the one-to-one correspondence of all projective representations of
$G$ and all linear representations of its covering group $G^*$.

Thus is the summary for this section. The uninterested reader
may now freely proceed to Section 3.
\subsection{The Covering Group}
Recall that a {\bf projective representation} of $G$ over
$\IC$ is a mapping $\rho : G \rightarrow GL(V)$ such that 
$\rho(\II_G) = \II_V$
and $\rho(x) \rho(y) = \alpha(x,y) \rho(x y)$
for any elements $x,y
\in G$. The function $\alpha$, known as the {\em cocycle}, is a map 
$G \times G \rightarrow \IC^*$ which is classified by $H^2(G,\IC^*)$,
the second $\IC^*$-valued cohomology of $G$.
This case of $\alpha=1$ trivially is of course our familiar ordinary
(non-projective) representation, which will be called {\bf linear}.

The Abelian group $H^2(G,\IC^*)$ is known as the {\bf Schur
Multiplier} of $G$ and will be
denoted by $M(G)$. Its triviality or otherwise is a discriminant of
whether $G$ admits projective representation. In a physical context,
knowledge of $M(G)$ provides immediate information as to the
possibility of turning on discrete torsion in the orbifold model under
study. A classification of $M(G)$ for all discrete finite 
subgroups of $SU(3)$ and the exceptional subgroups of $SU(4)$
was given in the companion work \cite{FHHP}. 

The study of the projective representations of a given group $G$
is greatly facilitated by introducing an auxilliary object $G^*$, the
{\bf covering group} of $G$, which ``lifts projective representations
to linear ones.'' Let us refresh our memory what this means.
Let there be a {\bf central extension} according to the exact sequence
$1 \rightarrow A \rightarrow G^* \rightarrow G \rightarrow 1$ such
that $A$ is in the centre of $G^*$. Thus we have $G^*/A \cong G$. Now
we say 
\begin{definition}
A projective representation $\rho$ of $G$ {\bf lifts} to a linear
representation $\rho^*$ of $G^*$ if\\ 
(i) $\rho^*(a \in A)$ is proportional to $\II$ and \\
(ii) there is a section\footnote{i.e., for the
        projection $f:G^*\rightarrow G$, $\mu \circ f = \II_G$.}
$\mu : G \rightarrow G^*$ such that $\rho(g) = \rho^*(\mu(g)),~
\forall g \in G$.
\end{definition}
Likewise it {\em lifts projectively} if $\rho(g) =
t(g) \rho^*(\mu(g))$ for a map (not necessarily a homomorphism)
$t:G \rightarrow \IC^*$.
\begin{definition}\label{defcover}
$G^*$ is called a {\bf covering group} (or otherwise 
known as the {\bf representation group}, Darstellungsgruppe)
of $G$ over $\IC$ if the following are satisfied:
\begin{description}
\item[(i)] $\exists$ a central extension $1 \rightarrow A \rightarrow G^*
\rightarrow G \rightarrow 1$ such that any projective representation
of $G$ lifts projectively to an ordinary representation of $G^*$;
\item[(ii)] $|A| = |M(G)| = |H^2(G,\IC^*)|$.
\end{description}
\end{definition}

The covering group will play a central r\^{o}le in our work; as we will
show in a moment, {\em the matter content of an orbifold theory
with group $G$ having discrete torsion switched-on is
encoded in the quiver diagram of $G^*$}.

For actual computational purposes, the following theorem, initially
due to Schur, is of extreme importance:
\begin{theorem}\label{cover}{\rm (\cite{Karp} p143)}
$G^{\star}$ is a covering
group of $G$ over $\IC$ if and only if the following conditions hold:\\
(i) $G^{\star}$ has a finite subgroup $A$ with $A\subseteq
	Z(G^{\star}) \cap [G^{\star},G^{\star}]$;\\
(ii) $G \cong G^{\star}/A$;\\
(iii) $|A|=|M(G)|$.
\end{theorem}
In the above, $[G^{\star},G^{\star}]$ is the {\bf derived group}
$G^{*'}$ of $G^*$. We remind ourselves that for a
group $H$, $H' := [H,H]$ is the group generated by elements of
the form $[x,y] := xyx^{-1}y^{-1}$ for $x,y \in H$.
We stress that conditions (ii) and (iii) are easily satisfied while
(i) is the more stringent imposition.

The solution of the problem of finding covering groups
is certainly {\em not} unique:
$G$ in general may have more than one covering groups (e.g.,
the quaternion and the dihedral group of order 8 are both covering
groups of $\IZ_2 \times \IZ_2$). The problem of finding the necessary
conditions which two groups $G_1$ and $G_2$ must satisfy in order for
both to be covering groups of the same group $G$ is a classical one.

The well-known solution starts with the following
\begin{definition}
\label{isoclinic}
Two groups $G$ and $H$ are said to be {\bf isoclinic} if there exist two 
isomorphisms
\[
\alpha : G/Z(G) \stackrel{\cong}{\rightarrow} H/Z(H) \quad {\rm and} \quad
\beta : G' \stackrel{\cong}{\rightarrow} H'
\]
such that
$\alpha(x_1 Z(G))=x_2 Z(H)~~{\rm and}~~\alpha(y_1 Z(G))=y_2 Z(H)
\Rightarrow \beta([x_1,y_1])= [x_2,y_2],$
\end{definition}
where we have used the standard notation that $x Z(G)$ is a coset
representative in $G/Z(G)$.
We note in passing that every Abelian group is obviously 
isoclinic to the trivial group $\gen{\II}$.
%To be able to characterise pairs of isoclinic groups we need one
%additional concept:
%\begin{definition}
%If $G$ and $H$ are groups and $f : G \rightarrow H$ a surjective
%homomorphism, we call $f$ isoclinic if 
%\[
%{\rm Ker}(f) \cap G' = 1.
%\]
%\end{definition}
%Then we have the following useful lemma:
%\begin{lemma}{\rm (\cite{Karp} p439)}
%Two groups $G$ and $H$ are isoclinic if there is an isoclinic
%surjective homomorphism $f : G \rightarrow H$.
%\end{lemma}

We introduce this concept of isoclinism because of the following
important Theorem of Hall:
\begin{theorem}{\rm (\cite{Karp} p441)}\label{Hall}
Any two covering groups of a given finite group $G$ are isoclinic.
\end{theorem}

Knowing that the covering groups of $G$ are not isomorphic
to each other, but isoclinic, a natural question to ask is how many 
non-isomorphic covering groups can one have. Here a theorem due to
Schur shall be useful:
\begin{theorem}{\rm (\cite{Karp} p149)}
\label{num_cover}
For a finite group $G$, let
\[
G/G' = \IZ_{e_1}\times ...\times \IZ_{e_r}
\]
and
\[
M(G)= \IZ_{f_1}\times ...\times \IZ_{f_s}
\]
be decompositions of these Abelian groups into cyclic factors. 
Then the number of 
non-isomorphic covering groups of $G$ is less than or 
equal to
\[
\prod_{1\leq i\leq r,1\leq j \leq s} \gcd(e_i,f_j).
\]
\end{theorem}
\subsection{Projective Characters}
With the preparatory remarks in the previous subsection, we now delve
headlong into the heart of the matter. By virtue of the construction
of the covering group $G^*$ of $G$ , we have the following 1-1
correpondence which will enable us to compute $\alpha$-projective
representations of $G$ in terms of the linear representations of
$G^*$:
\begin{theorem}{\sf [Theorema Egregium]} 
{\rm (\cite{Karp} p139; \cite{Hoff-Hum} p8)}
Let $G^*$ be the covering group of $G$ and $\lambda :
A \rightarrow \IC^*$ a homomorphism. Then
\begin{description}
\item[(i)] For every linear representation $L : G^* \rightarrow GL(V)$ 
of $G^*$ such that $L(a)=\lambda(a) \II_V~\forall a \in A$, 
there is an induced projective representation $P$ on $G$ defined by
\[
P(g) := L(r(g)), ~\forall~ g \in G,
\]
with $ r : G \rightarrow G^*$ the map that associates to each coset $g \in
G \cong G^*/A $ a representative element\footnote{i.e., $r(g)A
	\rightarrow g$  is the isomorphism $G^*/A
	\stackrel{\cong}{\rightarrow} G$.} in $G^*$; and
vice versa,
\item[(ii)] Every $\alpha$-projective representation for $\alpha \in
M(G)$ lifts to an ordinary representation of $G^*$.
\end{description}
\end{theorem}
An immediate consequence of the above is the fact that knowing the
linear characters of $G^*$ suffices to establish
the projective characters of $G$ for all $\alpha$ \cite{Humph}. 
This should ease
our initial fear in that {\em one does not need to know a priori the
specific values of the cocycles $\alpha(x,y)$ for all $x,y \in G$ (a
stupendous task indeed) in order to construct the $\alpha$-projective
character table for $G$.}

We shall leave the uses of this crucial observation to later
discussions. For now, let us focus on some explicit computations of
covering groups.
\section{Explicit Calculation of Covering Groups}
To theory we must supplant examples and to abstraction, concreteness.
We have prepared ourselves in the previous section the rudiments of
the theory of covering groups; in the present section we will
demonstrate these covers for the discrete finite subgroups of
$SU(3)$. First we shall illustrate our techniques with the case of
$D_{2n}$, the ordinary dihedral group, before tabulating the complete
results.
\subsection{The Covering Group of The Ordinary Dihedral Group}
The presentation of the ordinary dihedral group of order $2n$ is
standard (the notation is different from some of our earlier papers
(e.g. \cite{ZD}) where the following would be called $D_n$):
\[
D_{2n}= \gen{\tilde{\alpha},\tilde{\beta} | \tilde{\alpha}^n=1, \tilde{\beta}^2=1, 
	\tilde{\beta} \tilde{\alpha} \tilde{\beta}^{-1} = \tilde{\alpha}^{-1}}.
\]
We recall from \cite{FHHP} that the Schur Multiplier for $G = D_{2n}$ is
$\IZ_2$ when $n$ is even and trivial otherwise, thus we restrict
ourselves only to the case of $n$ even.
We let $M(D_{2n})$ be $A = \IZ_2$ generated by $\{ a |
 a^2=\II\}$. We let the covering group be $G^* =
\gen{\alpha,\beta, a}$.

Now having defined the generators we proceed to constrain relations
thereamong. Of course, $A \subset Z(G^*)$ immediately implies that
$\alpha a= a \alpha$ and
$\beta a= a \beta$. Moreover, $\alpha,\beta$ must map
to $\tilde{\alpha},\tilde{\beta}$ when we identify $G^{\star}/A \cong
D_{2n}$  (by part (ii) of Theorem \ref{cover}). This means that $\II_G$
must have a preimage in $A \subset G^*$, giving us: $\alpha^n \in
A, \beta^2 \in A$ and $\beta \alpha \beta^{-1} \alpha \in A$ by virtue
of the presentation of $G$. And hence we have 8 possibilities, each
being a central extension of $D_{2n}$ by $A$:
\begin{equation}
\label{extension}
\ba{l}
G^*_1=\gen{\alpha,\beta, a |\alpha a= a\alpha,~~
\beta a= a\beta,~~  a^2=1,~~ \alpha^n=1,~~
\beta^2=1,~~ \beta \alpha \beta^{-1}= \alpha^{-1}}
\\
G^*_2=\gen{\alpha,\beta, a |\alpha a= a\alpha,~~
\beta a= a\beta,~~  a^2=1,~~ \alpha^n=1,~~
\beta^2=1,~~ \beta \alpha \beta^{-1}= \alpha^{-1} a}
\\
G^*_3=\gen{\alpha,\beta, a |\alpha a= a\alpha,~~
\beta a= a\beta,~~  a^2=1,~~ \alpha^n=1,~~
\beta^2= a,~~ \beta \alpha \beta^{-1}= \alpha^{-1}}
\\
G^*_4=\gen{\alpha,\beta, a |\alpha a= a\alpha,~~
\beta a= a\beta,~~  a^2=1,~~ \alpha^n=1,~~
\beta^2= a,~~ \beta \alpha \beta^{-1}= \alpha^{-1}
 a}
\\
G^*_5=\gen{\alpha,\beta, a |\alpha a= a\alpha,~~
\beta a= a\beta,~~  a^2=1,~~ \alpha^n= a,~~
\beta^2=1,~~ \beta \alpha \beta^{-1}= \alpha^{-1}}
\\
G^*_6=\gen{\alpha,\beta, a |\alpha a= a\alpha,~~
\beta a= a\beta,~~  a^2=1,~~ \alpha^n= a,~~
\beta^2=1,~~ \beta \alpha \beta^{-1}= \alpha^{-1} a}
\\
G^*_7=\gen{\alpha,\beta, a |\alpha a= a\alpha,~~
\beta a= a\beta,~~  a^2=1,~~ \alpha^n= a,~~
\beta^2= a,~~ \beta \alpha \beta^{-1}= \alpha^{-1}}
\\
G^*_8=\gen{\alpha,\beta, a |\alpha a= a\alpha,~~
\beta a= a\beta,~~  a^2=1,~~ \alpha^n= a,~~
\beta^2= a,~~ \beta \alpha \beta^{-1}= \alpha^{-1} a}
\ea
\end{equation}

Therefore, conditions (ii) and (iii) of Theorem \ref{cover} are
satified. One must check (i) to distinguish the covering group among
these 8 central extensions in \eref{extension}. Now since $A$ is
actually the centre, it suffices to check whether $A \subset
G^{*'}_i = [G^*_i,G^*_i]$.

We observe $G^*_1$ to be precisely $D_{2n}\times \IZ_2$, from which we
have $G^{*'}_1 = \IZ_{n/2}$, generated by $\alpha^2$. 
Because $A = \{\II,a\}$ clearly is not included in this $\IZ_{n/2}$
we conclude that $G^*_1$ is not the covering group. 
For $G^*_2$, we have $G^{*'}_2 = \gen{\alpha^2 a}$, which means that
when $n/2=\odd$ (recall that $n=\even$), $G^{*'}_2$ can contain $a$
and hence $A \subset G^{*'}_2$, whereby making $G^*_2$ a covering group. 
By the same token we find that
$G^{*'}_3 = \gen{\alpha^2}$,
$G^{*'}_4 = \gen{\alpha^2 a}$,
$G^{*'}_5 = \gen{\alpha^2}$,
$G^{*'}_6 = \gen{\alpha^2 a}$, and
$G^{*'}_7 = \gen{\alpha^2}$.
We summarise these results in the following table:
\[
\ba{c|c|c|c|c}
{\rm Group}  & G^{*'} & Z(G^*) & G^*/Z(G^*) & \mbox{Covering Group?}\\ \hline
G^*_1  & \IZ_{n/2}=\gen{\alpha^2 } & \IZ_2\times \IZ_2=\gen{a,\alpha^{n/2} }
& D_{n} & $no$ \\
G^*_2(n=4k+2) & \IZ_n=\gen{\alpha^2a } & \IZ_2=\gen{ a } & D_{2n} &
$yes$ \\
G^*_2(n=4k)  & \IZ_{n/2}=\gen{ \alpha^2a } & 
\IZ_2\times \IZ_2=\gen{ a,\alpha^{n/2} }& D_{n}& $no$ \\
G^*_3  & \IZ_{n/2}=\gen{ \alpha^2} & \IZ_2\times
\IZ_2=\gen{a,\alpha^{n/2} }& D_{n} & $no$ \\
G^*_4(n=4k+2)  & \IZ_n=\gen{ \alpha^2a} & \IZ_2=\gen{ a }& D_{2n} & $yes$
\\
G^*_4(n=4k)  & \IZ_{n/2}=\gen{ \alpha^2a} & \IZ_2\times \IZ_2=\gen
{ a,\alpha^{n/2} } & D_{n} & $no$ \\
G^*_5  & \IZ_n=\gen{\alpha^2 } &  \IZ_2=\gen{ a }& D_{2n} & $yes$ \\
G^*_6(n=4k+2)  & \IZ_{n/2}=\gen{\alpha^2a} & \IZ_4=\gen{\alpha^{n/2} }
& D_{n} & $no$ \\
G^*_6(n=4k)  & \IZ_n=\gen{\alpha^2a} & \IZ_2=\gen{ a } & D_{2n} & $yes$\\
G^*_7  & \IZ_n=\gen{\alpha^2 } &  \IZ_2=\gen{ a }& D_{2n} & $yes$\\
G^*_8(n=4k+2)  & \IZ_{n/2}=\gen{ \alpha^2a }&  \IZ_4=\gen{\alpha^{n/2} }
 & D_{n}  & $no$ \\
G^*_8(n=4k)  & \IZ_n=\gen{ \alpha^2a }&  \IZ_2=\gen{ a }& D_{2n} & $yes$\\
\ea
\]
Whence we see that $G^*_1$ and $G^*_3$ are not
covering groups, while for $n/2 = \odd$ $G^*_{2,4}$ are covers, for
$n/2 = \even$ $G^*_{6,8}$ are covers as well 
and finally $G^*_{5,7}$ are always
covers. Incidentally, $G^*_7$ is actually the binary dihedral group and
we know that it is indeed the (double) covering group from \cite{FHHP}.
Of course in accordance with Theorem \ref{Hall}, these different
covers must be isoclinic to each other. Checking against Definition
\ref{isoclinic}, we see that for $G^*$ being $G^*_{2,4}$ with $n=4k+2$,
$G^*_{6,8}$ with $n = 4k$ and $G^*_{5,7}$ for all even $n$, $G^{*'} \cong
\IZ_n$ and $G^*/Z(G^*) \cong D_{2n}$; furthermore the isomorphisms
$\alpha$ and $\beta$ in the Definition are easily seen to satisfy the
prescribed conditions. Therefore all these groups are indeed isoclinic.
We make one further remark, for both the 
cases of $n=4k$ and $n=4k+2$, we have found 4
non-isomorphic covering groups. Recall Theorem \ref{num_cover},
here we have $f_1=2$ and $G/G'=\IZ_2\times \IZ_2$ (note that
$n$ is even) and so $e_1=e_2=2$, whence the upper limit is exactly
$2\times 2=4$ which is saturated here. This demonstrates that our method
is general enough to find all possible covering groups.
\newpage
\subsection{Covering Groups for the Discrete Finite Subgroups of
	$SU(3)$}
By methods entirely analogous to the one presented in the above
subsection, we can arrive at the covering groups for the discrete
finite groups of $SU(3)$ as tabulated in \cite{FHHP}. 
Let us list the results (of course in comparison with Table 3.2
in \cite{FHHP}, those with trivial Schur Multipliers have no
covering groups and will not be included here).
Of course, as mentioned earlier, the covering group is not unique. The
particular ones we have chosen in the following table are the same as
generated by the computer package GAP using the Holt algorithm \cite{GAP}.
\vspace{1.0in}
\subsection*{Intransitives}
\beq \label{ZZ}
\hspace{-1.8in}
\ba{ll}
\bullet \quad G = & \IZ_m \times \IZ_n = 
	\gen{\tilde{\alpha},\tilde{\beta} | 
	\tilde{\alpha}^n=1, \tilde{\beta}^m=1,
	\tilde{\alpha}\tilde{\beta}=\tilde{\beta}\tilde{\alpha}};\\
	& M(G) = \IZ_{p=\gcd(m,n)} = \gen{a | a^p=\II}; \\
	&  G^* = \gen{\alpha, \beta,  a |\alpha  a= a\alpha,
	\beta a=   a\beta,    a^p=1, \alpha^n=1,
	\beta^m=1, \alpha \beta =\beta \alpha  a}\\
\ea
\eeq

\beq \label{ZbinD}
\hspace{-0.3in}
\ba{ll}
\bullet \quad  G = & \gen{\IZ_{n=4k} , \widehat{D_{2m}}} = 
	\gen{\tilde{\alpha},\tilde{\beta},\tilde{\gamma}|
	\tilde{\alpha}\tilde{\beta}=\tilde{\beta}\tilde{\alpha},
	\tilde{\alpha}\tilde{\gamma}=\tilde{\gamma} \tilde{\alpha},
	\tilde{\alpha}^{n/2}=\tilde{\beta}^m, 
	\tilde{\beta}^{2m}=1, \tilde{\beta}^m=\tilde{\gamma}^2,
	\tilde{\gamma}\tilde{\beta}\tilde{\gamma}^{-1}=\tilde{\beta}^{-1}};\\
	& \left\{
	\ba{ll}
	m~\even \quad & M(G) = \IZ_2 \times \IZ_2=\gen{a,b|a^2=1=b^2, ab=ba};\\
		& \ba{ll}
		  G^* = & \langle 
			\alpha, \beta, \gamma,a,b| ab=ba,
			\alpha a=a \alpha, \alpha b=b \alpha, 
			\beta a= a \beta, \beta b= b \beta,\\
			& \gamma a=a \gamma, \gamma b=b \gamma, a^2=1=b^2,
				\alpha\beta =\beta\alpha a, 
				\alpha \gamma=\gamma  \alpha b,\\
			& \alpha^{n/2}=\beta^m, \beta^{2m}=1, 
				\beta^m=\gamma^2, \gamma \beta
				\gamma^{-1}=\beta^{-1} \rangle
		 \ea \\
	m~\odd,  \quad & M(G) = \IZ_2=\gen{a | a^2=1 };\\
		& \ba{ll}
		G^* = & \langle
		\alpha, \beta, \gamma,a | a^2=1,\alpha a=a \alpha,
		\beta a= a \beta,\gamma a=a \gamma, \alpha\beta
		=\beta\alpha, \\ 
		& \alpha \gamma=\gamma  \alpha a, 
		 \alpha^{n/2}=\beta^m, \beta^{2m}=1, \beta^m=\gamma^2, 
		\gamma \beta \gamma^{-1}=\beta^{-1} \rangle
		\ea \\
	%m~\odd, \gcd(n,4) \ne 1 \quad & M(G) = \IZ_4=\gen{a | a^4=1}; \\
	%	& \ba{ll}
	%	G^* = & \langle \alpha, \beta, \gamma,a | a^4=1,\alpha a=a \alpha,
	%		\beta a= a \beta,\gamma a=a \gamma,
	%		\alpha^n=1, \\ 
	%		& \alpha\beta =\beta\alpha a^2,
	%		\alpha \gamma=\gamma  \alpha a,
	%		\beta^{2m}=1, \beta^m=\gamma^2, 
	%		\gamma \beta \gamma^{-1}=\beta^{-1} \rangle
	%	\ea \\
	\ea
	\right.
\ea
\eeq

\beq \label{ZE6}
\hspace{-0.1in}
\ba{lll}
\bullet \quad G = & \gen{\IZ_{n=3k},\widehat{E_6}} & \\
& k~\odd
&
G \cong \IZ_n \times \widehat{E_6} = 
\gen{\tilde{\alpha},\tilde{\beta},\tilde{\gamma}|
\tilde{\alpha}\tilde{\beta}=\tilde{\beta}\tilde{\alpha},
\tilde{\alpha}\tilde{\gamma}=\tilde{\gamma} \tilde{\alpha},
\tilde{\alpha}^n=1, \tilde{\beta}^3=\tilde{\gamma}^3=
(\tilde{\beta}\tilde{\gamma})^2}; \\
& & M(G) = \IZ_3 = \gen{a | a^3=\II}; \\
& & \ba{ll}
  G^* = & \langle
\alpha, \beta, \gamma,a | a^3=1,\alpha a=a \alpha,
\beta a= a \beta,\gamma a=a \gamma, \alpha^n=1, \\ 
& \alpha\beta =\beta\alpha a^{-1}, \alpha \gamma=\gamma  \alpha a,
\beta^3=\gamma^3=(\beta\gamma)^2 \rangle
\ea \\
& k = 2(2p+1) & G \cong \IZ_{n/2} \times \widehat{E_6} \\
& k = 4p & G \cong (\IZ_{n} \times \widehat{E_6})/\IZ_2 =
\gen{\tilde{\alpha},\tilde{\beta},\tilde{\gamma}|
\tilde{\alpha}\tilde{\beta}=\tilde{\beta}\tilde{\alpha},
\tilde{\alpha}\tilde{\gamma}=\tilde{\gamma} \tilde{\alpha},
\tilde{\alpha}^{n/2}=\tilde{\beta}^3, \tilde{\beta}^3=\tilde{\gamma}^3=
(\tilde{\beta}\tilde{\gamma})^2}; \\
& & M(G) = \IZ_3 = \gen{a | a^3=\II}; \\
& & \ba{ll}
  G^* = & \langle
\alpha, \beta, \gamma,a | a^3=1,\alpha a=a \alpha,
\beta a= a \beta,\gamma a=a \gamma, \alpha^{n/2}=\beta^3, \\ 
& \alpha\beta =\beta\alpha a^{-1}, \alpha \gamma=\gamma  \alpha a,
\beta^3=\gamma^3=(\beta\gamma)^2 \rangle
\ea \\
\ea\eeq
\newpage
\beq \label{ZE7}
\hspace{-1.9in}
\ba{ll}
\bullet \quad G = & \gen{\IZ_{n=4k},\widehat{E_7}} =
\gen{
\tilde{\alpha},\tilde{\beta},\tilde{\gamma}|
\tilde{\alpha}\tilde{\beta}=\tilde{\beta}\tilde{\alpha},
\tilde{\alpha}\tilde{\gamma}=\tilde{\gamma} \tilde{\alpha},
\tilde{\alpha}^{n/2}=\tilde{\beta}^4, \tilde{\beta}^4=\tilde{\gamma}^3=
(\tilde{\beta}\tilde{\gamma})^2
};\\
& M(G) = \IZ_2 = \gen{a | a^2=\II}; \\
& \ba{ll}
G^* = & \langle \alpha, \beta, \gamma,a |  a^2=1,\alpha a=a \alpha,
\beta a= a \beta,\gamma a=a \gamma, \alpha^{n/2}=\beta^4, \\ 
& \alpha \beta =\beta \alpha a, \alpha \gamma=\gamma \alpha, 
\beta^4=\gamma^3 = (\beta \gamma)^2  \rangle
\ea \ea \vspace{-1.0in} \eeq
\beq \label{ZD}
\hspace{-0.3in}
\ba{lll}
\bullet \quad  G = & \gen{\IZ_n,  D_{2m}} & \\  
& 
n~\odd, m~\even  & G = \IZ_n \times   D_{2m}=
\langle
\tilde{\alpha},\tilde{\beta},\tilde{\gamma}| \tilde{\alpha}^n
=1,
\tilde{\alpha}\tilde{\beta}=\tilde{\beta}\tilde{\alpha},
\tilde{\alpha}\tilde{\gamma}=\tilde{\gamma}\tilde{\alpha},
\tilde{\beta}^m=1, \\
&&\qquad \qquad
\tilde{\gamma}^2=1, 
\tilde{\gamma} \tilde{\beta}\tilde{\gamma}^{-1}=\tilde{\beta}^{-1}
\rangle;\\
&  & M(G) = \IZ_2 = \gen{a|a^2= 1};\\
& &\ba{ll}
G^*= & \langle \alpha, \beta, \gamma,a |  a^2=1,
a (\alpha/\beta/\gamma)=(\alpha/\beta/\gamma) a, 
\alpha (\beta/\gamma)=(\beta/\gamma) \alpha, \alpha^n=1,
\\ & \beta^m=a, \gamma^2=1, \gamma \beta \gamma^{-1}=\beta^{-1} \rangle
\ea \\
& n~\even,m~\odd & G = \IZ_n \times   D_{2m} \\
& & M(G) = \IZ_2 = \gen{a|a^2= 1};\\
& & \ba{ll}
G^* = & \langle \alpha, \beta, \gamma,a |  a^2=1,
a (\alpha/\beta/\gamma)=(\alpha/\beta/\gamma) a, 
\alpha \beta=\beta \alpha,\alpha \gamma =\gamma \alpha a, \alpha^n=1,
\\ & \beta^m=1, \gamma^2=1, \gamma \beta \gamma^{-1}=\beta^{-1}
\rangle \ea \\
& m~\even, n=2(2l+1) & G = \IZ_{n/2} \times   D_{2m} \\
& n=4k, m=2(2l+1)  & G = (\IZ_{n} \times   D_{2m})/\IZ_2=
\langle
\tilde{\alpha},\tilde{\beta},\tilde{\gamma}| \tilde{\alpha}^{n/2}
=\beta^{m/2},
\tilde{\alpha}\tilde{\beta}=\tilde{\beta}\tilde{\alpha},
\tilde{\alpha}\tilde{\gamma}=\tilde{\gamma}\tilde{\alpha},\\
&&\qquad \qquad
\tilde{\beta}^m=1, \tilde{\gamma}^2=1, 
\tilde{\gamma} \tilde{\beta}\tilde{\gamma}^{-1}=\tilde{\beta}^{-1}
\rangle;\\
&  & M(G) = \IZ_2 = \gen{a|a^2= 1};\\
& &\ba{ll}
G^*= & \langle \alpha, \beta, \gamma,a |  a^2=1,
a (\alpha/\beta/\gamma)=(\alpha/\beta/\gamma) a, 
\alpha \beta=\beta \alpha, \alpha \gamma=\gamma \alpha a,
\\ & \alpha^{n/2}=\beta^{m/2},
 \beta^m=1, \gamma^2=1, \gamma \beta \gamma^{-1}=\beta^{-1} \rangle
\ea \\
& n=4k, m=4l  & G = (\IZ_{n} \times   D_{2m})/\IZ_2 \\
&  & M(G) = \IZ_2 \times \IZ_2 = \gen{a,b|a^2= 1, b^2=1, ab=ba};\\
& &\ba{ll}
G^*= & \langle \alpha, \beta, \gamma,a,b |  a^2=1,
a (\alpha/\beta/\gamma)=(\alpha/\beta/\gamma) a, 
\alpha \beta=\beta \alpha b, \alpha \gamma=\gamma \alpha a,
\\ & \alpha^{n/2}=\beta^{m/2},
 \beta^m=1, \gamma^2=1, \gamma \beta \gamma^{-1}=\beta^{-1} \rangle
\ea \\
%n~\even,m~\even \quad & M(G) = \IZ_2 \times \IZ_2 \times \IZ_2 = 
%	\gen{a,b,c|a^2=b^2=c^2=1};\\
%& \ba{ll}
%G^*= & \langle
%\alpha, \beta, \gamma,a, b,c |  a^2=1,b^2=1, c^2=1,
%(a/b/c) (\alpha/\beta/\gamma/a/b/c)= \\
%& = (\alpha/\beta/\gamma/a/b/c) (a/b/c),\\ 
%&\alpha \beta=\beta \alpha a, \alpha \gamma =\gamma \alpha b, \alpha^n=1,
% \beta^m=1, \gamma^2=c, \gamma \beta \gamma^{-1}=\beta^{-1}\rangle
%\ea \\
\ea \eeq

where we have used the shorthand notation $(x/y/\ldots/z)$ to mean the
relation to be applied to each of the elements $x,y,\ldots,z$.
\subsection*{Transitives}
We first have the two infinite series.
\beq \label{del3}
\ba{ll}
\bullet \quad G = & \Delta(3n^2) = \gen{\alpha,\beta,\gamma |
	\alpha^n = \beta^n = \gamma^3 = 1,
        \alpha \beta = \beta \alpha,
        \alpha \gamma =  \gamma \alpha^{-1} \beta,
        \beta \gamma \alpha = \gamma};\\
& \left\{ \ba{ll}
\gcd(n,3) = 1, n~\even \quad & M(G) = \IZ_n = \gen{a | a^n=1};\\
	& \ba{ll} G^* = & \langle
	\alpha, \beta, \gamma, a | 
	(\alpha/\beta/\gamma) a = a(\alpha/\beta/\gamma),\\
	&
         a^n = \alpha^n a^{n/2} = \beta^n a^{n/2} = \gamma^3 = 1,\\
	&
        \alpha \beta = \beta \alpha a,
        \alpha \gamma = \gamma \alpha^{-1} \beta,
        \beta \gamma \alpha = \gamma
		\rangle; \ea \\
\gcd(n,3) = 1, n~\odd \quad & M(G) = \IZ_n;\\
	& \ba{ll} G^* = & \langle
	\alpha, \beta, \gamma, a | 
	(\alpha/\beta/\gamma) a = a(\alpha/\beta/\gamma),\\
	&
         a^n = \alpha^n = \beta^n = \gamma^3 = 1,\\
	&
        \alpha \beta = \beta \alpha a,
        \alpha \gamma = \gamma \alpha^{-1} \beta,
        \beta \gamma \alpha = \gamma
		\rangle; \ea \\
\gcd(n,3) \ne 1, n~\even \quad & M(G) = \IZ_n \times \IZ_3 = 
			\gen{a,b|a^n=1,b^3=1};\\
	& \ba{ll} G^* = & \langle
	\alpha, \beta, \gamma, a, b | 
	(\alpha/\beta/\gamma)(a/b) = (a/b)(\alpha/\beta/\gamma),\\
	&
        ab=ba, a^n=b^3 = \gamma^3 =\alpha^n a^{n/2} b = 1, \\
	&
        \beta^n a^{n/2}= b,
        \alpha \beta = \beta \alpha a b,
        \alpha \gamma = \gamma \alpha^{-1} \beta,
        \beta \gamma \alpha = \gamma
		\rangle; \ea \\
\gcd(n,3) \ne 1, n~\odd \quad & M(G) = \IZ_n \times \IZ_3; \\
	& \ba{ll} G^* = & \langle
	\alpha, \beta, \gamma, a, b |
	(\alpha/\beta/\gamma)(a/b) = (a/b)(\alpha/\beta/\gamma),\\
	&
	a^n=b^3 = \gamma^3 =\alpha^n b = \beta^n b^{-1}= 1, \\
	&
	ab=ba,
        \alpha \beta = \beta \alpha a b,
        \alpha \gamma = \gamma \alpha^{-1} \beta,
        \beta \gamma \alpha = \gamma
		\rangle; \ea \\
\ea
\right.
\ea
\eeq

\beq \label{del6}
\ba{ll}
\bullet \quad G = & \Delta(6n^2) = 
	\langle \alpha,\beta,\gamma,\delta |
	\alpha^n = \beta^n = \gamma^3 = \delta^2 = 1,
        \alpha \beta = \beta \alpha,
        \alpha \gamma = \gamma \alpha^{-1} \beta,
        \beta \gamma \alpha = \gamma,\\
	&
	\qquad \qquad \qquad
	\alpha \delta \alpha = \delta,
        \beta \delta = \delta \alpha^{-1} \beta,
        \gamma \delta \gamma = \delta
	\rangle;\\
	& M(G) = \IZ_2 = \gen{a | a^2 = 1};\\
& \left\{ \ba{ll}
\ba{ll} \gcd(n,4) = 4 \quad G^* = & \langle
	\alpha, \beta, \gamma, \delta, a |
	\alpha^n = \beta^n = \gamma^3 = \delta^2 = a^2 = 1,\\
	&
        ~~(\alpha/\beta/\gamma/\delta) a = a (\alpha/\beta/\gamma/\delta),
	\alpha \beta = \beta \alpha a,
        \alpha \gamma = \gamma \alpha^{-1} \beta,
        \beta \gamma \alpha = \gamma,\\
	&
        ~~\alpha \delta \alpha = \delta,
        \beta \delta = \delta \alpha^{-1} \beta,
        \gamma \delta \gamma = \delta
	\rangle; \ea \\
\ba{ll} \gcd(n,4) = 2 \quad G^* = & \langle
	\alpha, \beta, \gamma, \delta, a |
	\alpha^n a = \beta^n a = \gamma^3 = \delta^2 = a^2 = 1,\\
	&
        ~~(\alpha/\beta/\gamma/\delta) a = a (\alpha/\beta/\gamma/\delta),
	\alpha \beta = \beta \alpha a,
        \alpha \gamma = \gamma \alpha^{-1} \beta,
        \beta \gamma \alpha = \gamma,\\
	&
        ~~\alpha \delta \alpha = \delta,
        \beta \delta = \delta \alpha^{-1} \beta,
        \gamma \delta \gamma = \delta
	\rangle; \ea \\
\ea
\right.
\ea
\eeq
Next we present the three exceptionals that admit discrete torsion.
\beq \label{sig60}
\hspace{-2.2in}
\ba{ll}
\bullet \quad G = & \Sigma(60) \cong A_5 =
	\langle \alpha, \beta |
	\alpha^5 = \beta^3 = (\alpha \beta^{-1})^3 = (\alpha^2
	\beta)^2 = 1\\
	&
	\qquad \qquad \qquad \qquad \alpha \beta \alpha \beta \alpha \beta =  
	\alpha \gamma \alpha^{-1} \beta \alpha^2 \beta \alpha^{-2}
	\beta = 1
	\rangle;\\
	& M(G) = \IZ_2; \\
	& G^* = \langle \alpha, \beta, a|
	 \alpha^5 = a, \beta^3 = a^2 =1, (\alpha/\beta)a =
	a(\alpha/\beta)\\
	&
	\qquad \qquad (\alpha \beta^{-1})^3 = 1, (\alpha^2
	\beta)^2 = a
	\rangle; \\

\ea
\eeq

\beq \label{sig168}
\hspace{-1.0in}
\ba{ll}
\bullet \quad G = & \Sigma(168) = \gen{ \alpha, \beta, \gamma |
	\gamma^2 = \beta^3 = \beta \gamma \beta \gamma =
	(\alpha\gamma)^4 = 1,
	\alpha^2 \beta  = \beta \alpha, 
	\alpha^3 \gamma \alpha^{-1} \beta = \gamma \alpha \gamma 
	};\\
	& M(G) = \IZ_2; \\
	& G^*  = \langle \alpha, \beta, \gamma, \delta |
	\delta^2 = \gamma^2 \delta = \beta^3 \delta = (\beta \alpha)^3
	= (\alpha \gamma)^3 = 1, \\
	& \qquad \qquad
	\beta \gamma \beta = \gamma, \alpha \delta = \delta \alpha, 
	\beta^2 \alpha^2 \beta = \alpha, 
	\beta^{-1} \alpha^{-1} \beta \gamma \alpha^{-1} \gamma =
	\gamma \alpha \beta
	\rangle;\\
\ea
\eeq

\beq \label{sig1080}
\hspace{-1.7in}
\ba{ll}
\bullet \quad G = & \Sigma(1080) = \langle \alpha, \beta, \gamma, \delta |
	\alpha^5 = \beta^2 = \gamma^2 = \delta^2 = (\alpha \beta)^2 
	(\beta \gamma)^2 = (\beta \delta)^2 = 1,\\
	& \qquad \qquad \qquad \qquad 
	(\alpha \gamma)^3 = (\alpha \delta)^3 = 1, 
	\gamma \beta = \delta \gamma \delta, 
	\alpha^2 \gamma \beta \alpha^2 = \gamma \alpha^2 \gamma
	\rangle; \\
	& M(G) = \IZ_2; \\
	& G^* = \langle \alpha, \beta, \gamma, \delta, \epsilon |
	\alpha^5 = \epsilon^2 = \gamma^2 \epsilon^{-1} = \beta^2 \epsilon^{-1} 
	= \delta^2 \epsilon^{-1} = (\alpha \delta)^3 = 1, \\
	& \qquad \qquad \qquad \qquad
	\alpha^{-1} \epsilon \alpha = \beta^{-1} \epsilon \beta
	= \gamma^{-1} \epsilon \gamma
	= \delta^{-1} \epsilon \delta = \epsilon, \\
	& \qquad \qquad \qquad \qquad
	(\alpha \beta)^2 = (\beta \gamma)^2 = (\beta \delta)^2 =
	\gamma \beta \delta \gamma \delta =
	(\alpha \gamma)^3 = \epsilon,\\
	& \qquad \qquad \qquad \qquad
	\alpha^2 \gamma \beta \alpha^2 \gamma \alpha^{-2} \gamma=1
	\rangle;\\
\ea
\eeq
\section{Covering Groups, Discrete Torsion and Quiver Diagrams}
\subsection{The Method}
The introduction of the host of the above concepts is not without a
cause. In this section we shall provide an {\bf algorithm} which
permits the construction of the quiver $Q_{\alpha}(G,{\cal R})$ of an
orbifold theory with group $G$ having discrete torsion $\alpha$
turned-on, and with a linear representation ${\cal R}$ of $G$ acting 
on the transverse space. 

Our method dispenses of the need of the knowledge of the cocycles
$\alpha(x,y)$, which in general is a formidable task from the
viewpoint of cohomology, but which the
current literature may lead the reader to believe to be required for
finding the projective representations. We shall demonstrate that the
problem of finding these $\alpha$-representations is reducible to the far
more manageable duty of finding the covering group, constructing its character
table (which is of course straightforward) and then applying the usual
prodecure of extracting the quiver therefrom.
One advantage of this method is that we immediately obtain 
the quiver for all cocycles (including the trivial cocycle
which corresponds to having no discrete torsion at all) and in fact
the values of $\alpha(x,y)$ (which we shall address in the next
section) in a unified framework. 

All the data we require are\\
(i) $G$ and its (ordinary) character table char$(G)$;\\
(ii) The covering group $G^*$ of $G$ and its (ordinary) character
	table char$(G^*)$.

We first recall from \cite{Doug} that turning on discrete torsion
$\alpha$ in an orbifold projection amounts to using an
$\alpha$-projective representation\footnote{More rigorously, this
	statement holds only when the D-brane probe is pointlike in the
        orbifold directions. More generally, when D-brane probes
	extend along the orbifold directions, one may need to use 
	ordinary as well as projective 
	representations. For further details, please refer to
	\cite{Gab} as well as \cite{Craps1}.} 
 $\Gamma_{\alpha}$ of $g \in G$
\beq
\label{proj}
\Gamma_{\alpha}(g) \cdot A \cdot \Gamma^{-1}_{\alpha}(g) = A,
\qquad
\Gamma_{\alpha}(g) \cdot \Phi \cdot \Gamma^{-1}_{\alpha}(g) 
= {\cal R}(g) \cdot \Phi
\eeq
on the gauge field $A$ and matter fields $\Phi$.

The above equations have been phrased in a more axiomatic setting (in
the language of \cite{LNV}), by virtue of the fact that much of
ordinary representation theory of finite group extends in direct
analogy to the projective case, in \cite{AspinPles}.
{\em However, we hereby emphasize that with the aid of the
linear representation of the covering
group, one can perform orbifold projection with discrete torsion
entirely in the setting of \cite{LNV} without usage of the
formulae in \cite{AspinPles} generalised to twisted group algebras and
modules.}
In other words, if we use the matrix of the linear representation of $G^*$
instead of that of the corresponding projective representation of $G$, we will
arrive at the same gauge group and matter contents in the orbifold
theory. This can be alternatively shown as follows.

When we lift the
projective matrix representation of $G$ into the linear one of
$G^*$, every matrix $\rho(g)$ will 
map to $\rho(ga_i)$ for every $a_i\in A$. The crucial 
fact is that $\rho(ga_i)=\lambda_i \rho(g)$ 
where $\lambda_i$ is simply a phase factor. 
Now in \eref{proj} (cf. Sections 4.2 and 5 for more details),
$\Gamma_{\alpha}(g)$ and $\Gamma^{-1}_{\alpha}(g)$
always appear in pairs, when we replace them by
$\Gamma(ga_i)$ and $\Gamma^{-1}(ga_i)$,
the phase factor $\lambda_i$ will cancel out and leave the
result invariant. This shows that the two results,
the one given by projective matrix representations of $G$
and the other by linear matrix representations of $G^*$,
will give identical answers in orbifold projections.
\subsection{An Illustrative Example: $\Delta(3\times3^2)$}
Without much further ado, an illustrative example of the group
$\Delta(3\times3^2) \in SU(3)$ shall serve to enlighten the reader. We
recall from \eref{del3} that this group of order 27 has presentation
$\gen{\alpha,\beta,\gamma | \alpha^3 = \beta^3 = \gamma^3 = 1,
\alpha \beta = \beta \alpha,\alpha \gamma = \gamma \alpha^{-1} \beta,\beta
\gamma \alpha = \gamma}$ and its covering group of order 243 (since
the Schur Multiplier is $\IZ_3 \times \IZ_3$) is $G^* = \gen{\alpha,
\beta, \gamma, a, b | (\alpha/\beta/\gamma)(a/b) =
(a/b)(\alpha/\beta/\gamma), a^3=b^3 = \gamma^3 =\alpha^3 b = 
\beta^3 b^{-1}= 1,
ab=ba, \alpha \beta = \beta \alpha a b, \alpha \gamma = \gamma 
\alpha^{-1} \beta,
\beta \gamma \alpha = \gamma}$.

Next we require the two (ordinary) character tables. As pointed out in
the Nomenclatures section, character tables are given as $(r+1) \times
r$ matrices with $r$ being the number of conjugacy classes (and equivalently
the number of irreps), and the first row giving the conjugacy class numbers.
\beq \label{del3n3}
{\rm char}(\Delta(3\times3^2)) = 
{\tiny
\ba{|c|c|c|c|c|c|c|c|c|c|c|}
\hline
1 & 1 & 1 & 3 & 3 & 3 & 3 & 3 & 3 & 3 & 3 \\ \hline 1 & 1 & 1 & 1 & 1 & 1 & 1 & 1 & 1 & 1 & 1 \\ \hline 1 & 1 & 1 & 1 & 1 & \omega_3 & \omega_3 & \omega_3 & {\bar{\omega}_3} & 
   {\bar{\omega}_3} & {\bar{\omega}_3} \\ \hline 1 & 1 & 1 & 1 & 1 & {\bar{\omega}_3} & {\bar{\omega}_3} & {\bar{\omega}_3} & \omega_3 & \omega_3 & \omega_3 \\ \hline 1 & 1 & 1 & \omega_3 & {\bar{\omega}_3
   } & 1 & \omega_3 & {\bar{\omega}_3} & 1 & \omega_3 & {\bar{\omega}_3} \\ \hline 1 & 1 & 1 & \omega_3 & {\bar{\omega}_3} & \omega_3 & {\bar{\omega}_3} & 1 & {\bar{\omega}_3
   } & 1 & \omega_3 \\ \hline 1 & 1 & 1 & \omega_3 & {\bar{\omega}_3} & {\bar{\omega}_3} & 1 & \omega_3 & \omega_3 & {\bar{\omega}_3} & 1 \\ \hline 1 & 1 & 1 & {\bar{\omega}_3} & \omega_3 & 1 & {\bar{\omega}_3
   } & \omega_3 & 1 & {\bar{\omega}_3} & \omega_3 \\ \hline 1 & 1 & 1 & {\bar{\omega}_3} & \omega_3 & \omega_3 & 1 & {\bar{\omega}_3} & {\bar{\omega}_3} & \omega_3 & 1 \\ \hline 1 & 1 & 1 & {\bar{\omega}_3
   } & \omega_3 & {\bar{\omega}_3} & \omega_3 & 1 & \omega_3 & 1 & {\bar{\omega}_3} \\ \hline 3 & 3{\bar{\omega}_3} & 3\omega_3 & 0 & 0 & 0 & 0 & 0 & 0 & 0 & 0 \\ \hline 3 & 3\omega_3 & 3
   {\bar{\omega}_3} & 0 & 0 & 0 & 0 & 0 & 0 & 0 & 0 \\ \hline
\ea
}
\eeq
\beq \label{cgdel3n3}
\ba{l}
{\rm char}(\Delta(3\times3^2)^*) = \\
{\tiny
\ba{|c|c|c|c|c|c|c|c|c|c|c|c|c|c|c|c|c|c|c|c|c|c|c|c|c|c|c|c|c|c|c|c|c|c|c|}
\hline
   1 & 1 & 1 & 1 & 1 & 1 & 1 & 1 & 1 & 9 & 9 & 9 & 9 & 9 & 9 & 9 & 9 & 9 & 9 & 9 & 9 & 9 & 9 & 9 & 9 & 9 & 9 & 9 & 9 & 9 & 9 & 9 & 9 & 9 & 9 \\ \hline 1 & 1 & 1 & 1 & 
   1 & 1 & 1 & 1 & 1 & 1 & 1 & 1 & 1 & 1 & 1 & 1 & 1 & 1 & 1 & 1 & 1 & 1 & 1 & 1 & 1 & 1 & 1 & 1 & 1 & 1 & 1 & 1 & 1 & 1 & 1 \\ \hline 1 & 1 & 1 & 1 & 1 & 1 & 1 & 1 & 
   1 & 1 & 1 & 1 & 1 & 1 & 1 & 1 & 1 & \omega_3 & \omega_3 & \omega_3 & \omega_3 & \omega_3 & \omega_3 & \omega_3 & \omega_3 & \omega_3 & {\bar{\omega}_3} & {\bar{\omega}_3} & {\bar{\omega}_3} & {\bar{\omega}_3} & {\bar{\omega}_3
   } & {\bar{\omega}_3} & {\bar{\omega}_3} & {\bar{\omega}_3} & {\bar{\omega}_3} \\ \hline 1 & 1 & 1 & 1 & 1 & 1 & 1 & 1 & 1 & 1 & 1 & 1 & 1 & 1 & 1 & 1 & 1 & {\bar{\omega}_3
   } & {\bar{\omega}_3} & {\bar{\omega}_3} & {\bar{\omega}_3} & {\bar{\omega}_3} & {\bar{\omega}_3} & {\bar{\omega}_3} & {\bar{\omega}_3} & {\bar{\omega}_3
   } & \omega_3 & \omega_3 & \omega_3 & \omega_3 & \omega_3 & \omega_3 & \omega_3 & \omega_3 & \omega_3 \\ \hline 1 & 1 & 1 & 1 & 1 & 1 & 1 & 1 & 1 & 1 & 1 & \omega_3 & \omega_3 & \omega_3 & {\bar{\omega}_3} & {\bar{\omega}_3} & {\bar{\omega}_3
   } & 1 & 1 & 1 & \omega_3 & \omega_3 & \omega_3 & {\bar{\omega}_3} & {\bar{\omega}_3} & {\bar{\omega}_3} & 1 & 1 & 1 & \omega_3 & \omega_3 & \omega_3 & {\bar{\omega}_3} & {\bar{\omega}_3} & {\bar{\omega}_3
   } \\ \hline 1 & 1 & 1 & 1 & 1 & 1 & 1 & 1 & 1 & 1 & 1 & \omega_3 & \omega_3 & \omega_3 & {\bar{\omega}_3} & {\bar{\omega}_3} & {\bar{\omega}_3} & \omega_3 & \omega_3 & \omega_3 & {\bar{\omega}_3} & {\bar{\omega}_3
   } & {\bar{\omega}_3} & 1 & 1 & 1 & {\bar{\omega}_3} & {\bar{\omega}_3} & {\bar{\omega}_3
   } & 1 & 1 & 1 & \omega_3 & \omega_3 & \omega_3 \\ \hline 1 & 1 & 1 & 1 & 1 & 1 & 1 & 1 & 1 & 1 & 1 & \omega_3 & \omega_3 & \omega_3 & {\bar{\omega}_3} & {\bar{\omega}_3} & {\bar{\omega}_3} & {\bar{\omega}_3} & 
   {\bar{\omega}_3} & {\bar{\omega}_3} & 1 & 1 & 1 & \omega_3 & \omega_3 & \omega_3 & \omega_3 & \omega_3 & \omega_3 & {\bar{\omega}_3} & {\bar{\omega}_3} & {\bar{\omega}_3
   } & 1 & 1 & 1 \\ \hline 1 & 1 & 1 & 1 & 1 & 1 & 1 & 1 & 1 & 1 & 1 & {\bar{\omega}_3} & {\bar{\omega}_3} & {\bar{\omega}_3} & \omega_3 & \omega_3 & \omega_3 & 1 & 1 & 1 & {\bar{\omega}_3} & 
   {\bar{\omega}_3} & {\bar{\omega}_3} & \omega_3 & \omega_3 & \omega_3 & 1 & 1 & 1 & {\bar{\omega}_3} & {\bar{\omega}_3} & {\bar{\omega}_3
   } & \omega_3 & \omega_3 & \omega_3 \\ \hline 1 & 1 & 1 & 1 & 1 & 1 & 1 & 1 & 1 & 1 & 1 & {\bar{\omega}_3} & {\bar{\omega}_3} & {\bar{\omega}_3} & \omega_3 & \omega_3 & \omega_3 & \omega_3 & \omega_3 & \omega_3 & 1 & 1 & 1 & \bar{\omega}_3 & {\bar{\omega}_3} & {\bar{\omega}_3} & {\bar{\omega}_3} & {\bar{\omega}_3} & {\bar{\omega}_3
   } & \omega_3 & \omega_3 & \omega_3 & 1 & 1 & 1 \\ \hline 1 & 1 & 1 & 1 & 1 & 1 & 1 & 1 & 1 & 1 & 1 & {\bar{\omega}_3} & {\bar{\omega}_3} & {\bar{\omega}_3} & \omega_3 & \omega_3 & \omega_3 & {\bar{\omega}_3} & 
   {\bar{\omega}_3} & {\bar{\omega}_3} & \omega_3 & \omega_3 & \omega_3 & 1 & 1 & 1 & \omega_3 & \omega_3 & \omega_3 & 1 & 1 & 1 & {\bar{\omega}_3} & {\bar{\omega}_3} & {\bar{\omega}_3
   } \\ \hline 3 & 3 & 3 & 3 & 3 & 3 & 3 & 3 & 3 & 3{\bar{\omega}_3} & 3
   \omega_3 & 0 & 0 & 0 & 0 & 0 & 0 & 0 & 0 & 0 & 0 & 0 & 0 & 0 & 0 & 0 & 0 & 0 & 0 & 0 & 0 & 0 & 0 & 0 & 0 \\ \hline 3 & 3 & 3 & 3 & 3 & 3 & 3 & 3 & 3 & 3\omega_3 & 3
   {\bar{\omega}_3} & 0 & 0 & 0 & 0 & 0 & 0 & 0 & 0 & 0 & 0 & 0 & 0 & 0 & 0 & 0 & 0 & 0 & 0 & 0 & 0 & 0 & 0 & 0 & 0 \\ \hline 3 & 3\omega_3 & 3{\bar{\omega}_3} & 3 & 3
   \omega_3 & 3{\bar{\omega}_3} & 3 & 3\omega_3 & 3
   {\bar{\omega}_3} & 0 & 0 & 0 & 0 & 0 & 0 & 0 & 0 & 0 & 0 & 0 & X & Y & Z & 0 & 0 & 0 & 0 & 0 & 0 & 0 & 0 & 0 & P & M & N \\ \hline 3 & 3\omega_3 & 3
   {\bar{\omega}_3} & 3 & 3\omega_3 & 3{\bar{\omega}_3} & 3 & 3\omega_3 & 3
   {\bar{\omega}_3} & 0 & 0 & 0 & 0 & 0 & 0 & 0 & 0 & 0 & 0 & 0 & Z & X & Y & 0 & 0 & 0 & 0 & 0 & 0 & 0 & 0 & 0 & M & N & P \\ \hline 3 & 3\omega_3 & 3
   {\bar{\omega}_3} & 3 & 3\omega_3 & 3{\bar{\omega}_3} & 3 & 3\omega_3 & 3
   {\bar{\omega}_3} & 0 & 0 & 0 & 0 & 0 & 0 & 0 & 0 & 0 & 0 & 0 & Y & Z & X & 0 & 0 & 0 & 0 & 0 & 0 & 0 & 0 & 0 & N & P & M \\ \hline 3 & 3{\bar{\omega}_3} & 3
   \omega_3 & 3 & 3{\bar{\omega}_3} & 3\omega_3 & 3 & 3{\bar{\omega}_3} & 3
   \omega_3 & 0 & 0 & 0 & 0 & 0 & 0 & 0 & 0 & 0 & 0 & 0 & M & P & N & 0 & 0 & 0 & 0 & 0 & 0 & 0 & 0 & 0 & Z & Y & X \\ \hline 3 & 3{\bar{\omega}_3} & 3\omega_3 & 3 & 3
   {\bar{\omega}_3} & 3\omega_3 & 3 & 3{\bar{\omega}_3} & 3
   \omega_3 & 0 & 0 & 0 & 0 & 0 & 0 & 0 & 0 & 0 & 0 & 0 & N & M & P & 0 & 0 & 0 & 0 & 0 & 0 & 0 & 0 & 0 & Y & X & Z \\ \hline 3 & 3{\bar{\omega}_3} & 3\omega_3 & 3 & 3
   {\bar{\omega}_3} & 3\omega_3 & 3 & 3{\bar{\omega}_3} & 3
   \omega_3 & 0 & 0 & 0 & 0 & 0 & 0 & 0 & 0 & 0 & 0 & 0 & P & N & M & 0 & 0 & 0 & 0 & 0 & 0 & 0 & 0 & 0 & X & Z & Y \\ \hline 3 & 3 & 3 & 3\omega_3 & 3\omega_3 & 3\omega_3 & 3
   {\bar{\omega}_3} & 3{\bar{\omega}_3} & 3{\bar{\omega}_3} & 0 & 0 & A & -B & C & -A & 
    -C & B & 0 & 0 & 0 & 0 & 0 & 0 & 0 & 0 & 0 & 0 & 0 & 0 & 0 & 0 & 0 & 0 & 0 & 0 \\ \hline 3 & 3 & 3 & 3\omega_3 & 3\omega_3 & 3\omega_3 & 3{\bar{\omega}_3} & 3
   {\bar{\omega}_3} & 3{\bar{\omega}_3} & 0 & 0 & -B & C & A & -C & B & 
    -A & 0 & 0 & 0 & 0 & 0 & 0 & 0 & 0 & 0 & 0 & 0 & 0 & 0 & 0 & 0 & 0 & 0 & 0 \\ \hline 3 & 3 & 3 & 3\omega_3 & 3\omega_3 & 3\omega_3 & 3{\bar{\omega}_3} & 3{\bar{\omega}_3} & 3
   {\bar{\omega}_3} & 0 & 0 & C & A & -B & B & -A & -C & 0 & 0 & 0 & 0 & 0 & 0 & 0 & 0 & 0 & 0 & 0 & 0 & 0 & 0 & 0 & 0 & 0 & 0 \\ \hline 3 & 3\omega_3 & 3
   {\bar{\omega}_3} & 3\omega_3 & 3{\bar{\omega}_3} & 3 & 3{\bar{\omega}_3} & 3 & 3\omega_3 & 0 & 0 & 0 & 0 & 0 & 0 & 0 & 0 & -C & -A & B & 0 & 0 & 0 & 0 & 0 & 0 & 
    -B & C & A & 0 & 0 & 0 & 0 & 0 & 0 \\ \hline 3 & 3\omega_3 & 3{\bar{\omega}_3} & 3\omega_3 & 3{\bar{\omega}_3} & 3 & 3{\bar{\omega}_3} & 3 & 3
   \omega_3 & 0 & 0 & 0 & 0 & 0 & 0 & 0 & 0 & -A & B & -C & 0 & 0 & 0 & 0 & 0 & 0 & A & -B & C & 0 & 0 & 0 & 0 & 0 & 0 \\ \hline 3 & 3\omega_3 & 3{\bar{\omega}_3} & 3\omega_3 & 3
   {\bar{\omega}_3} & 3 & 3{\bar{\omega}_3} & 3 & 3\omega_3 & 0 & 0 & 0 & 0 & 0 & 0 & 0 & 0 & B & -C & -A & 0 & 0 & 0 & 0 & 0 & 0 & C & A & 
    -B & 0 & 0 & 0 & 0 & 0 & 0 \\ \hline 3 & 3{\bar{\omega}_3} & 3\omega_3 & 3\omega_3 & 3 & 3{\bar{\omega}_3} & 3{\bar{\omega}_3} & 3
   \omega_3 & 3 & 0 & 0 & 0 & 0 & 0 & 0 & 0 & 0 & 0 & 0 & 0 & 0 & 0 & 0 & M & N & P & 0 & 0 & 0 & Z & X & Y & 0 & 0 & 0 \\ \hline 3 & 3{\bar{\omega}_3} & 3\omega_3 & 3
   \omega_3 & 3 & 3{\bar{\omega}_3} & 3{\bar{\omega}_3} & 3
   \omega_3 & 3 & 0 & 0 & 0 & 0 & 0 & 0 & 0 & 0 & 0 & 0 & 0 & 0 & 0 & 0 & P & M & N & 0 & 0 & 0 & X & Y & Z & 0 & 0 & 0 \\ \hline 3 & 3{\bar{\omega}_3} & 3\omega_3 & 3
   \omega_3 & 3 & 3{\bar{\omega}_3} & 3{\bar{\omega}_3} & 3
   \omega_3 & 3 & 0 & 0 & 0 & 0 & 0 & 0 & 0 & 0 & 0 & 0 & 0 & 0 & 0 & 0 & N & P & M & 0 & 0 & 0 & Y & Z & X & 0 & 0 & 0 \\ \hline 3 & 3 & 3 & 3{\bar{\omega}_3} & 3
   {\bar{\omega}_3} & 3{\bar{\omega}_3} & 3\omega_3 & 3\omega_3 & 3\omega_3 & 0 & 0 & -A & -C & B & A & 
    -B & C & 0 & 0 & 0 & 0 & 0 & 0 & 0 & 0 & 0 & 0 & 0 & 0 & 0 & 0 & 0 & 0 & 0 & 0 \\ \hline 3 & 3 & 3 & 3{\bar{\omega}_3} & 3{\bar{\omega}_3} & 3{\bar{\omega}_3} & 3
   \omega_3 & 3\omega_3 & 3\omega_3 & 0 & 0 & -C & B & -A & -B & C & A & 0 & 0 & 0 & 0 & 0 & 0 & 0 & 0 & 0 & 0 & 0 & 0 & 0 & 0 & 0 & 0 & 0 & 0 \\ \hline 3 & 3 & 3 & 3
   {\bar{\omega}_3} & 3{\bar{\omega}_3} & 3{\bar{\omega}_3} & 3\omega_3 & 3\omega_3 & 3\omega_3 & 0 & 0 & B & -A & -C & C & A & 
    -B & 0 & 0 & 0 & 0 & 0 & 0 & 0 & 0 & 0 & 0 & 0 & 0 & 0 & 0 & 0 & 0 & 0 & 0 \\ \hline 3 & 3\omega_3 & 3{\bar{\omega}_3} & 3{\bar{\omega}_3} & 3 & 3\omega_3 & 3\omega_3 & 3
   {\bar{\omega}_3} & 3 & 0 & 0 & 0 & 0 & 0 & 0 & 0 & 0 & 0 & 0 & 0 & 0 & 0 & 0 & X & Z & Y & 0 & 0 & 0 & P & N & M & 0 & 0 & 0 \\ \hline 3 & 3\omega_3 & 3
   {\bar{\omega}_3} & 3{\bar{\omega}_3} & 3 & 3\omega_3 & 3\omega_3 & 3
   {\bar{\omega}_3} & 3 & 0 & 0 & 0 & 0 & 0 & 0 & 0 & 0 & 0 & 0 & 0 & 0 & 0 & 0 & Y & X & Z & 0 & 0 & 0 & N & M & P & 0 & 0 & 0 \\ \hline 3 & 3\omega_3 & 3
   {\bar{\omega}_3} & 3{\bar{\omega}_3} & 3 & 3\omega_3 & 3\omega_3 & 3
   {\bar{\omega}_3} & 3 & 0 & 0 & 0 & 0 & 0 & 0 & 0 & 0 & 0 & 0 & 0 & 0 & 0 & 0 & Z & Y & X & 0 & 0 & 0 & M & P & N & 0 & 0 & 0 \\ \hline 3 & 3{\bar{\omega}_3} & 3
   \omega_3 & 3{\bar{\omega}_3} & 3\omega_3 & 3 & 3\omega_3 & 3 & 3{\bar{\omega}_3} & 0 & 0 & 0 & 0 & 0 & 0 & 0 & 0 & -B & A & C & 0 & 0 & 0 & 0 & 0 & 0 & -C & B & 
    -A & 0 & 0 & 0 & 0 & 0 & 0 \\ \hline 3 & 3{\bar{\omega}_3} & 3\omega_3 & 3{\bar{\omega}_3} & 3\omega_3 & 3 & 3\omega_3 & 3 & 3
   {\bar{\omega}_3} & 0 & 0 & 0 & 0 & 0 & 0 & 0 & 0 & A & C & -B & 0 & 0 & 0 & 0 & 0 & 0 & -A & -C & B & 0 & 0 & 0 & 0 & 0 & 0 \\ \hline 3 & 3{\bar{\omega}_3} & 3
   \omega_3 & 3{\bar{\omega}_3} & 3\omega_3 & 3 & 3\omega_3 & 3 & 3{\bar{\omega}_3} & 0 & 0 & 0 & 0 & 0 & 0 & 0 & 0 & C & -B & A & 0 & 0 & 0 & 0 & 0 & 0 & B & -A & 
    -C & 0 & 0 & 0 & 0 & 0 & 0 \\ \hline
\ea
}\\
\ea
\eeq
with $A := -\omega_3 + \bar{\omega}_3, B := \omega_3 +
2\bar{\omega}_3, C := 2\omega_3 + \bar{\omega}_3; M := -\omega_9^2 - 2
\bar{\omega}_9^4, N := \omega_9^2 + \bar{\omega}_9^4, P := -\omega_9^2
+ \bar{\omega}_9^4; X := \omega_9^4 - \bar{\omega}_9^2, Y :=
\omega_9^4 + 2 \bar{\omega}_9^2, Z := -2 \omega_9^4 - \bar{\omega}_9^2$.

A comparative study of these two tables shall suffice to demonstrate
the method. We have taken extreme pains to re-arrange the columns and
rows of char$(G^*)$ for the sake of perspicuity; whence we immediately
observe that char$(G)$ and char$(G^*)$ are unrelated but that the
latter is organised in terms of ``cohorts'' \cite{Chat} of the
former. What this means is as follows: columns 1 through 9 of
char$(G^*)$ have their first 11 rows (not counting the row of class
numbers) identical to the first column of char$(G)$, so too is
column 10 of char$(G^*)$ with column 2 of char$(G)$, {\it et cetera}
with $\{11\} \rightarrow \{3\}$, $\{12,13,14\} \rightarrow \{4\}$, 
$\{15,16,17\} \rightarrow \{5\}$, $\{18,19,20\} \rightarrow \{6\}$, 
$\{21,22,23\} \rightarrow \{7\}$, $\{24,25,26\} \rightarrow \{8\}$, 
$\{27,28,29\} \rightarrow \{9\}$, $\{30,31,32\} \rightarrow \{10\}$, 
and $\{33,34,35\} \rightarrow \{11\}$; using the notation that $\{X\}
\rightarrow \{Y\}$ for the first 11 rows of columns $\{X\} \subset$
 char$(G^*)$ are mapped to column $\{Y\} \subset$ char$(G)$. These are
the so-called ``splitting conjugacy classes'' in $G^*$ which give the
(linear) char$(G)$ \cite{Hoff-Hum}.
In other words, (though the conjugacy class numbers may differ),
up to repetition char$(G) \subset$ char$(G^*)$. This of course is in the
spirit of the technique of Fr{\o}benius Induction of finding
the character table of a group from that of its subgroup; for a
discussion of this in the context of orbifolds, the reader
is referred to \cite{Step}. 
Thus the first 11 rows of char$(G^*)$ corresponds exactly to the
{\em linear irreps} of $G$. The rest of the rows
we shall shortly observe to correspond to the
projective representations. 

To understand these above remarks, 
let $A := \Z_3 \times \Z_3$ so that $G^* / A \cong G$ as
in the notation of Section 2. Now $A \subseteq Z(G^*)$, hence the
matrix forms of all of its elements must be $\lambda \II_{d\times d}$,
where $d$ is the dimension of the irreducible representation and 
$\lambda$ some phase factor. Indeed the first 9 columns of char$(G^*)$
have conjugacy class number 1 and hence correspond to elements of this
centre. Bearing this in mind, if we only tabulated the phases
$\lambda$ (by suppressing the factor $d=$ 1 or 3 coming from $\II_{d\times
d}$) of these first 9 columns, we arrive at the following table
(removing the first row of conjugacy class numbers):
\[
\ba{|c||c|c|c|c|c|c|c|c|c|}
\hline
\mbox{rows} & \II & a  & a^2 & b  & ab  & a^2 b & b^2 & a b^2 & a^2 b^2 \\ \hline
2 - 12	& 1   & 1    &1      & 1   & 1   & 1      & 1   & 1   & 1  \\ \hline
13-15	& 1   &\omega_3&\bar{\omega}_3  & 1   &\omega_3 &\bar{\omega}_3 
                     & 1   &\omega_3 &\bar{\omega}_3\\ \hline
16-18	& 1 &\bar{\omega}_3 & \omega_3 & 1 &\bar{\omega}_3 & \omega_3 
                       & 1 &\bar{\omega}_3 & \omega_3 \\ \hline
19-21	& 1   & 1    &1 &\omega_3&\omega_3&\omega_3  &\bar{\omega}_3&\bar{\omega}_3
                        &\bar{\omega}_3 \\ \hline
22-24	& 1   &\omega_3&\bar{\omega}_3 &\omega_3 &\bar{\omega}_3 &1 
                     &\bar{\omega}_3  & 1  &\omega_3 \\ \hline
25-27	& 1 &\bar{\omega}_3 & \omega_3  &\omega_3  & 1  &\bar{\omega}_3 
                     &\bar{\omega}_3   &\omega_3 & 1\\ \hline
28-30	& 1   & 1    &1  &\bar{\omega}_3&\bar{\omega}_3&\bar{\omega}_3
                        &\omega_3&\omega_3&\omega_3 \\ \hline
31-33	& 1   &\omega_3&\bar{\omega}_3  &\bar{\omega}_3  & 1  &\omega_3
                        &\omega_3 &\bar{\omega}_3 &1 \\ \hline
34-36	& 1 &\bar{\omega}_3 & \omega_3  &\bar{\omega}_3  &\omega_3& 1
                        &\omega_3 & 1 & \bar{\omega}_3 \\ \hline
\ea
\]
The astute reader would instantly recognise this to be the character
table of $\IZ_3 \times \IZ_3 = A$ (and with foresight we have labelled
the elements of the group in the above table). This certainly is to be
expected: $G^*$ can be written as cosets $gA$ for $g \in G$, whence
lifting the (projective) matrix representation $M(g)$ of $g$ simply
gives $\lambda M(g)$ for $\lambda$ a {\em phase factor} correponding
to the representation (or character as $A$ is always Abelian) of $A$.

What is happening should be clear: all of this is merely Part (i) of Theorem
\ref{cover} at work. The phases $\lambda$ are precisely as
described in the theorem. The trivial phase 1 gives rows $2-12$, or
simply the ordinary representation of $G$ while the remaining 8
non-trivial phases give, in groups of 3 rows from char$(G^*)$, the
projective representations of $G$. And to determine to which cocycle the
projective representation belongs, we need and only need to determine
the the 1-dimensional irreps of $A$.
We shall show in Section 5 how to read out the actual cocycle values
$\alpha(g,h)$ for $g,h\in G$
directly with the knowledge of $A$ and $G^*$ without char$(G^*)$.

Enough said on the character tables. Let us proceed to analyse the
quiver diagrams. Detailed discussions had already been presented in
the case of the dihedral group in \cite{FHHP}. Let us recapitulate the
key points. It is the group action on the Chan-Paton bundle that we
choose to be projective, the space-time action inherited from ${\cal
N}=4$ R-symmetry remain ordinary. In other words, ${\cal R}$ from
\eref{proj} must still be a linear representation.

Now we evoke an obvious though handy result: 
the tensor product of an $\alpha$-projective representation 
with that of a $\beta$-representation gives
an $\alpha\beta$-projective representation (cf. \cite{Karp} p119), i.e., 
\beq\label{ab}
\Gamma_\alpha(g) \otimes \Gamma_\beta(g) = \Gamma_{\alpha\beta}(g).
\eeq
We recall that from \eref{proj} and in the language of
\cite{AspinPles,LNV}, the bi-fundamental matter
content $a_{ij}^{\cal R}$ is given in terms of the irreducible representations $R_i$
of $G$ as
\beq
\label{proj2}
{\cal R} \otimes R_i = \bigoplus\limits_j a_{ij}^{\cal R} R_j,
\eeq
(with of course ${\cal R}$ linear and $R_i$ projective representations).
Because ${\cal R}$ is an $\alpha=1$ (linear) representation, \eref{ab}
dictates that if $R_i$ in \eref{proj2} is a $\beta$-representation,
then the righthand thereof must be written entirely in terms of 
$\beta$-representations $R_j$. In
other words, the various projective representations corresponding to
the different cocycles should not mix under \eref{proj2}.
What this signifies for the matter matrix is that $a_{ij}^{\cal R}$ is
block-diagonal and the quiver diagram $Q(G^*,{\cal R})$ for $G^*$ {\em splits} into
precisely $|A|$ pieces, one of which is the ordinary (linear) quiver
for $G$ and the rest, the various quivers each corresponding to a
different value of the cocycle.

Thus motivated, let us present the quiver diagram for
$\Delta(3\times3^2)^*$ in \fref{f:del3_3}. The splitting does indeed
occur as desired, into precisely $|\IZ_3 \times \IZ_3|=9$ pieces, with
(i) being the usual $\Delta(3\times3^2)$ quiver
(cf. \cite{HanHe,Muto}) and the rest, the quivers corresponding to
the 8 non-trivial projective representations.
\EPSFIGURE[ht]{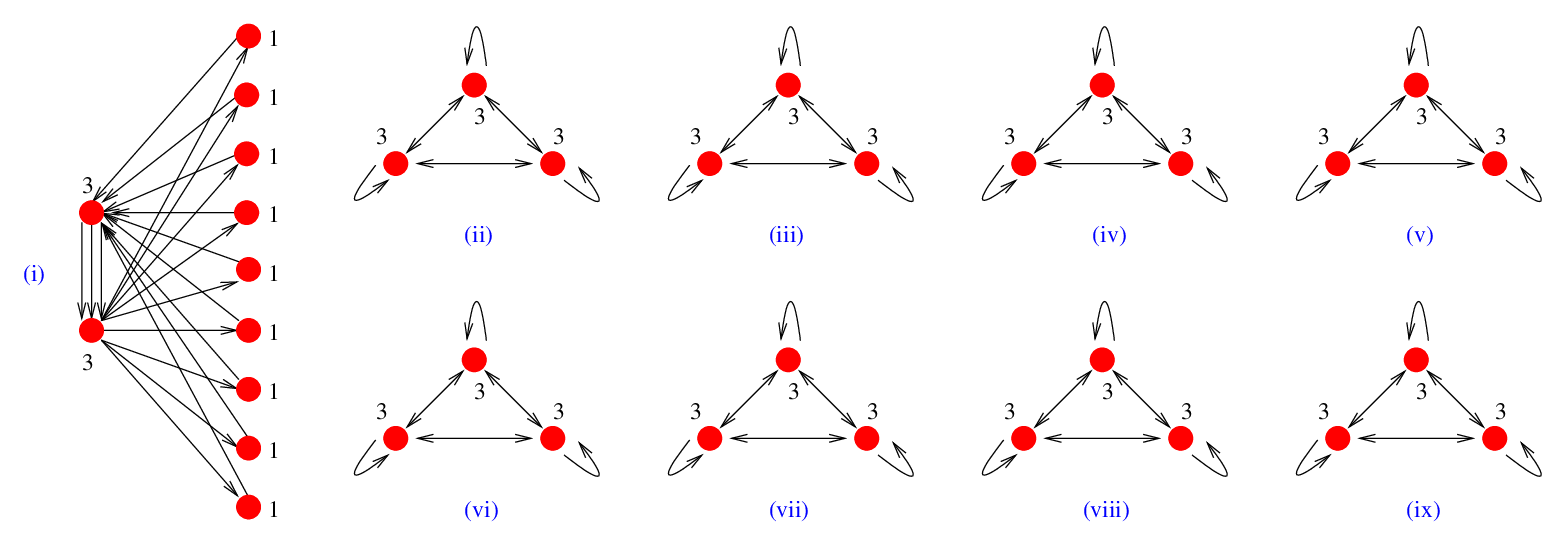,width=6in}
{
\label{f:del3_3}
The Quiver Diagram for $\Delta(3\times3^2)^*$ (the Space Invaders
Quiver): piece (i) corresponds to
the usual quiver for $\Delta(3\times3^2)$
while the remaining 8 pieces (ii) to (ix)
are for the cases of the 8 non-trivial discrete torsions (out of the
$\IZ_3 \times \IZ_3$) turned on.
}
\subsection{The General Method}
Having expounded upon the detailed example of $\Delta(3\times3^2)$ and
witnessed the subtleties, we now present, in an algorithmic manner,
the general method of computing the quiver diagram for an orbifold $G$
with discrete torsion turned on:
\begin{enumerate}
\item Compute the character table char$(G)$ of $G$;
\item Compute a covering group $G^*$ of $G$ and its character table 
	char$(G^*)$;
\item Judiciously re-order the rows and columns of char$(G^*)$:
	\begin{itemize}
	\item Columns must be arranged into cohorts of char$(G)$,
		i.e., group the columns which contain a corresponding
		column in char$(G)$ together;
	\item Rows must be arranged so that modulo the dimension of
		the irreps, the columns with conjugacy class number 1 
		must contain the character table of the Schur
		Multiplier $A = M(G)$ (recall that $G^*/A\cong G$);
	\item Thus char$(G)$ is a sub-matrix (up to repetition) of
		char$(G^*)$;
	\end{itemize}
\item Compute the (ordinary) matter matrix 
	$a_{ij}^{\cal R}$ and hence
	the quiver $Q(G^*,{\cal R})$ for a representation
	${\cal R}$ which corresponds to a linear representation of $G$.
\end{enumerate}
Now we have our final result:
\begin{theorem}
$Q(G^*,{\cal R})$
has $|M(G)|$ disconnected components (sub-quivers) in 1-1 correspondence with
the quivers $Q_{\alpha}(G,{\cal R})$ of $G$ for all possible cocycles
(discrete torsions) $\alpha \in A=M(G)$. Symbolically,
\[
Q(G^*,{\cal R}) = \bigsqcup_{\alpha \in A} Q_{\alpha}(G,{\cal R}).
\]
\end{theorem}
In particular, $Q(G^*,{\cal R})$ contains a piece for the trivial
$\alpha =1$ which is precisely the case without discrete torsion,
viz., $Q(G,{\cal R})$.

This algorithm facilitates enormously the investigation of the
matter spectrum of orbifold gauge theories with discrete torsion as
the associated quivers can be found without any recourse to explicit
evaluation of the cocycles and projective character tables.
Another fine feature of this new understanding is that, not only the 
matter content, but also the superpotential can be directly calculated 
by the explicit formulae in \cite{LNV} using the ordinary
Clebsch-Gordan coefficients of $G^*$. 

A remark is at hand. We have mentioned in Section 2 that the covering
group $G^*$ is not unique. How could we guarantee that the quivers
obtained at the end of the day will be independent of the choice of
the covering group? We appeal directly to the discussion in the
concluding paragraph of Subsection 4.1, where we remarked that using
the explicit form of \eref{proj}, we see that the phase factor
$\lambda$ (being a $\IC$-number) always cancels out. In other
words, the linear representation of whichever $G^*$ we use,
when applied to orbifold projections \eref{proj} shall result in the
same matrix form for the projective representations of $G$.
Whence we conclude that the
quiver $Q(G^*,{\cal R})$ obtained at the end will {\it ipso facto} be
independent of the choice of the covering group $G^*$.
\\ \\ \\ \\
\subsection{A Myriad of Examples}
With the method at hand, we move on to the host of other subgroups of
$SU(3)$ as tabulated in \cite{FHHP}.
The character tables char$(G)$ and char$(G^*)$ will be left
to the appendix lest the reader be too distracted.
We present the cases of $\Sigma(60,168,1080)$, the exceptionals which
admit nontrivial discrete torsion and some first members of the Delta
series in \fref{f:sig60} to \fref{f:del3_5}.
\EPSFIGURE[h]{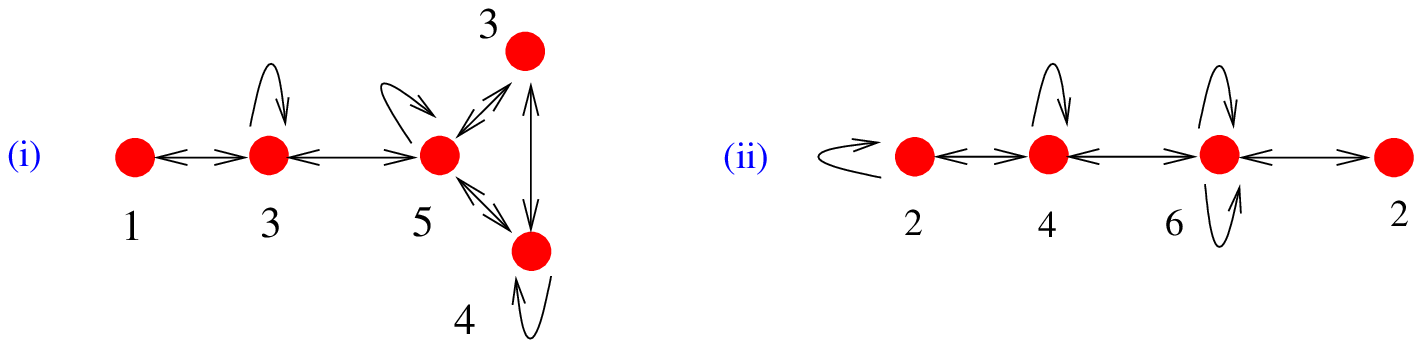,width=4.5in}
{\label{f:sig60}
The quiver diagram of $\Sigma(60)^*$: piece (i) is the ordinary quiver
of $\Sigma(60)$ and piece (ii) has discrete torsion turned on.
}
\EPSFIGURE[h]{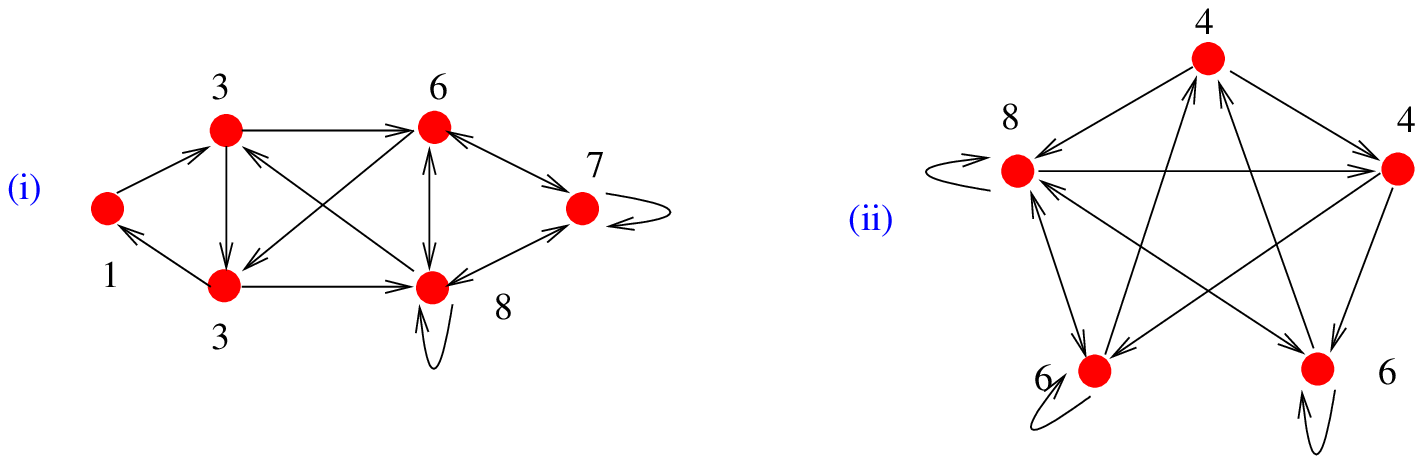,width=4.5in}
{\label{f:sig168}
The quiver diagram of $\Sigma(168)$: piece (i) is the ordinary quiver
of $\Sigma(168)$ and piece (ii) has discrete torsion turned on.
}
{\vspace{-1.0cm}}
\EPSFIGURE[ht]{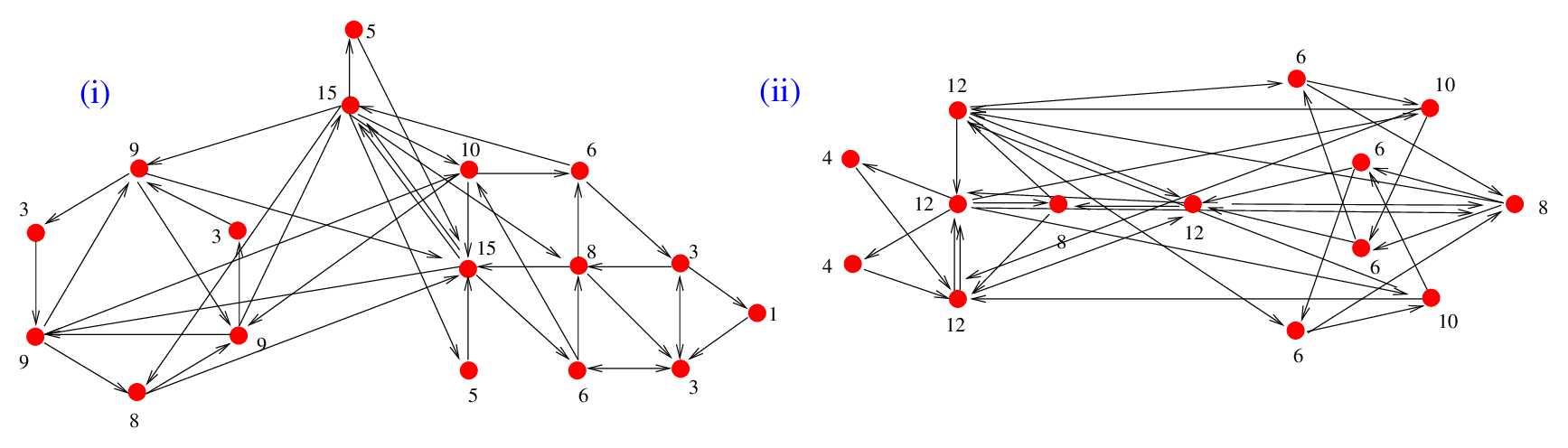,width=7.5in}
{\label{f:sig1080}
The quiver diagram of $\Sigma(1080)$: piece (i) is the ordinary quiver
of $\Sigma(1080)$ and piece (ii) has discrete torsion turned on.
}
\newpage
\EPSFIGURE[ht]{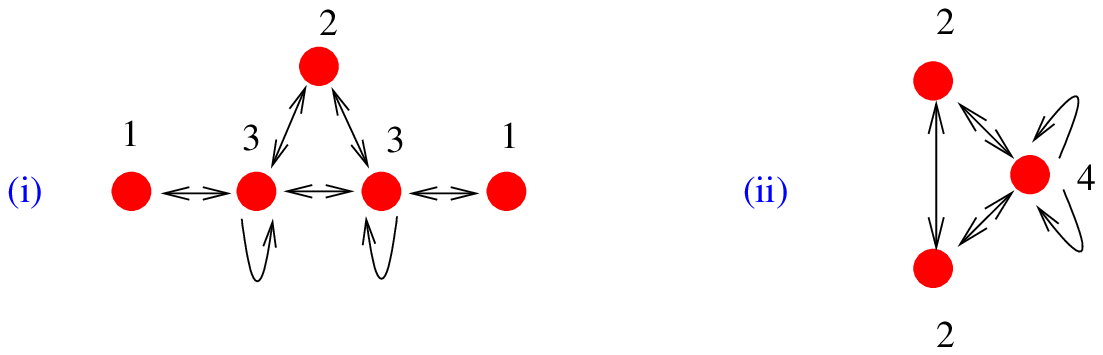,width=4.2in}
{\label{f:del6_2}
The quiver diagram of $\Delta(6\times 2^2)$: piece (i) is the ordinary quiver
of $\Delta(6\times 2^2)$ and piece (ii) has discrete torsion turned on.
}
\EPSFIGURE[ht]{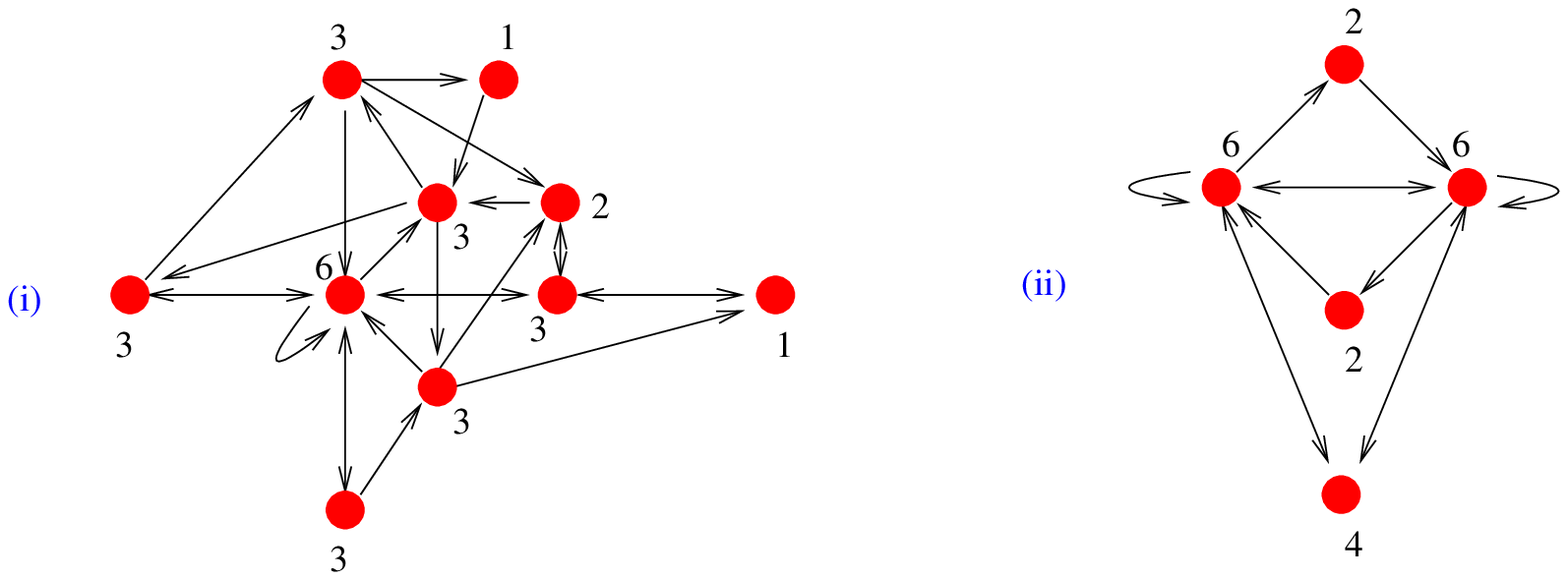,width=4.5in}
{\label{f:del6_4}
The quiver diagram of $\Delta(6\times 4^2)$: piece (i) is the ordinary quiver
of $\Delta(6\times 4^2)$ and piece (ii) has discrete torsion turned on.
}
\EPSFIGURE[ht]{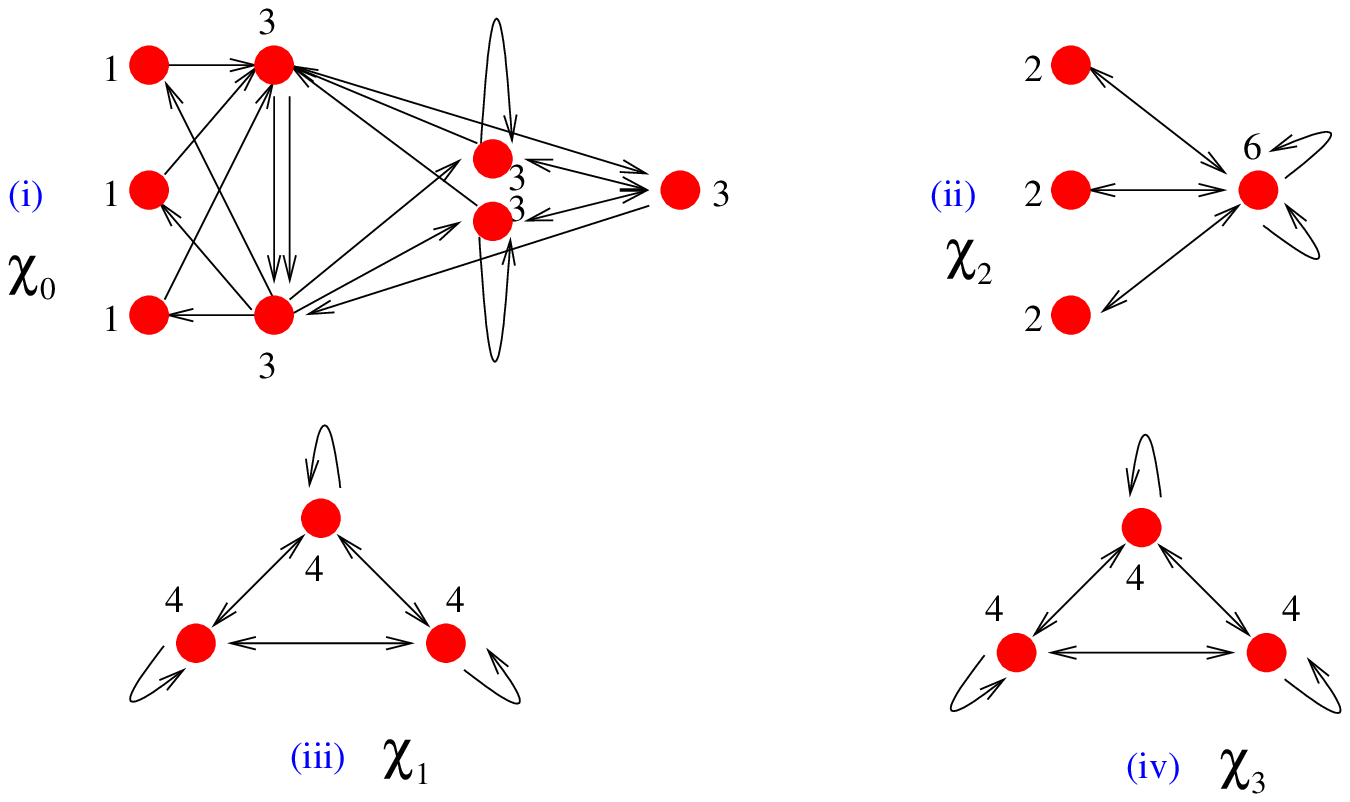,width=4.5in}
{\label{f:del3_4}
The quiver diagram of $\Delta(3\times 4^2)$: piece (i) is the ordinary quiver
of $\Delta(3\times 4^2)$ and pieces (ii-iv) have discrete torsion
turned on. We recall that the Schur Multiplier is $\IZ_4$.
}
\EPSFIGURE[ht]{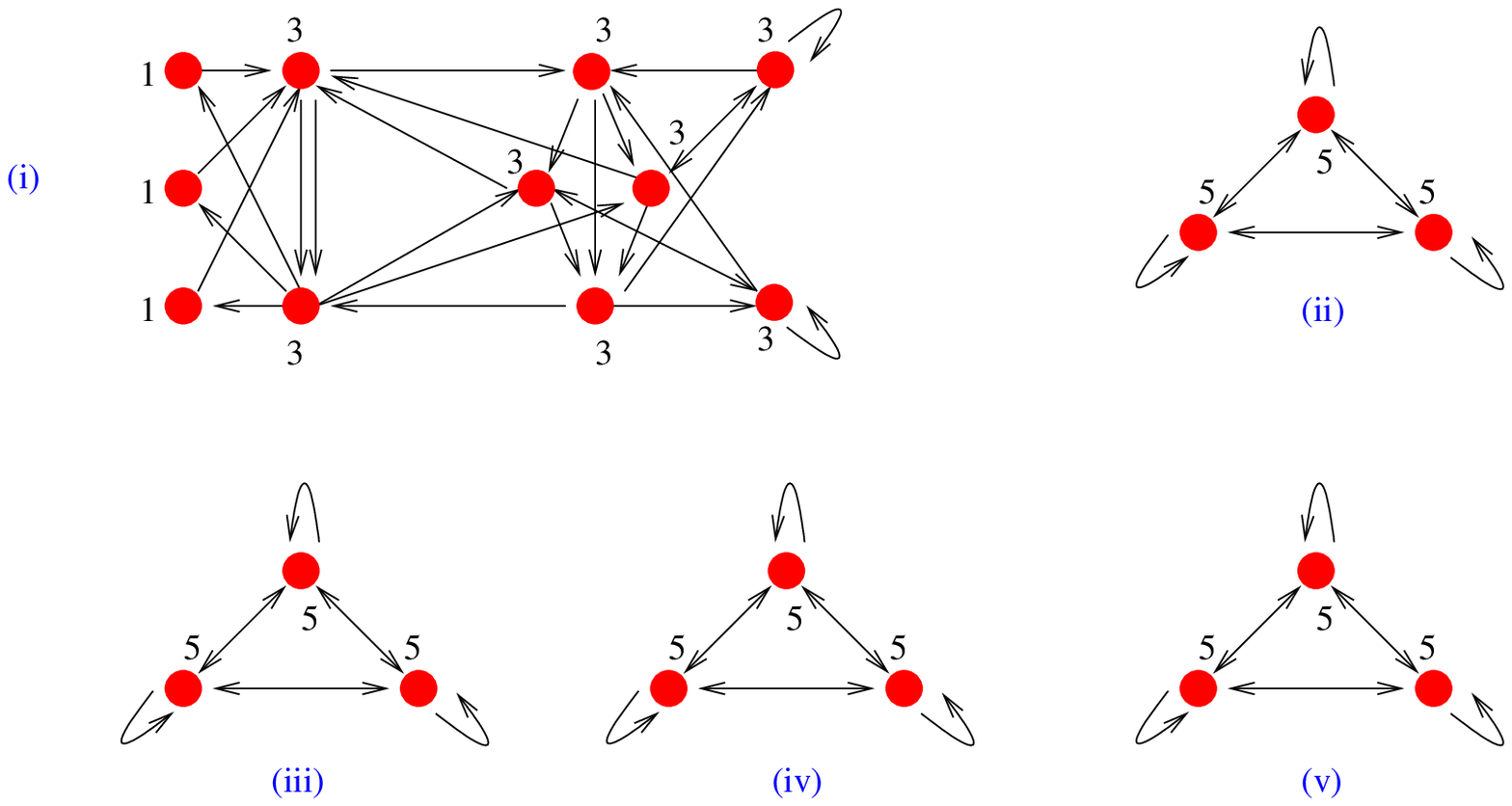,width=4.5in}
{\label{f:del3_5}
The quiver diagram of $\Delta(3\times 5^2)$: piece (i) is the ordinary quiver
of $\Delta(3\times 4^2)$ and pieces (ii-v) have discrete torsion
turned on. We recall that the Schur Multiplier is $\IZ_5$.
}
\section{Finding the Cocycle Values}
As advertised earlier, a useful by-product of the method is that we
can actually find the values of the 2-cocycles from the covering group.
Here we require even less information: only $G^*$ and not even
char$(G^*)$ is needed.

Let us recall some facts from Subsection 4.2.
The Schur multiplier is $A \subset Z(G^*)$, so every element therein has
its own conjugacy class in $G^*$. 
Hence for all linear representations of $G^*$, the character of $a_k
\in A$ will have the form $d \chi_i(a_k)$ where $d$ is the dimension
of that particular irrep of $G^*$ and $\chi_i(a_k)$ is the character
of ${a_k}$ in 
$A$ in its $i-th$ 1-dimensional irrep ($A$ is always Abelian and thus has
only 1-dimensional irreps).
This property has a very important consequence: merely reading out
the factor $\chi_i(a_k)$ from char$(G^*)$, we can
determine which linear representations will give which {\em projective}
representations of $G$. 
Indeed, two projective representations of $G$ belong to the same cocycle
{\em when and only when} the factor 
$\chi_i(a_k)$ is the same for every $a_k \in A$. 

Next we recall how to construct the matrix forms of projective
representations of $G$. 
$G^{\star}/A\equiv G$ implies that $G^*$ can be decomposed into cosets
$\bigcup\limits_{g \in G} g A$. Let $g a_i \in G^*$ correspond
canonically to $\tilde{g} \in G$ for some fixed $a_i \in A$; then
the matrix form of $\tilde{g}$ can be set to that of $g a_i$ and
furnishes the projective representation of $\tilde{g}$.
Different choices of $a_i$ will give different but projectively
equivalent projective representations of $G$.

Note that if we have
$\tilde{g_i}\tilde{g_j}=\tilde{g_k}$ in $G$, then in $G^*$, $g_i g_j= g_k
a_{ij}^{k}$, or
$(g_i a_i)(g_j a_j) = g_k a_k (a_{ij}^k a_i a_j a_k^{-1})$, but since
$(g_i a_i)$ is the projective matrix form for $\tilde{g_i} \in G$,
this is exactly the definition of the cocyle from which we read:
\beq
\label{cocycle}
\alpha(\tilde{g_i},\tilde{g_j})= \chi_p (a_{ij}^k a_i a_j a_k^{-1}),
\eeq
where $\chi_p(a)$ is the $p$-th character of the linear representation of
$a \in A$ defined above.

We can prove that \eref{cocycle} satisfies the 2-cocycle axioms (i)
and (ii).
Firstly notice that if $\tilde{g_i}= \II \in G$, 
we have $g_i = \II \in G^{\star}$; whence $a_{ij}^{k}= \delta_j^k~\forall~i$ and 
\[
(i)\quad \alpha(\II,\tilde{g_j})=\chi_p(\delta_j^k a_j a_k^{-1}) = \chi_p(\II)=1.
\]
Secondly if we assume that $\tilde{g_i}\tilde{g_j}=\tilde{g_q}$,
$\tilde{g_q}\tilde{g_k}=\tilde{g_h}$ and $\tilde{g_j}\tilde{g_k}=\tilde{g_l}$,
we have
$\alpha(\tilde{g_i},\tilde{g_j})\alpha(\tilde{g_i}\tilde{g_j},\tilde{g_k})
= \chi_p (a_{ij}^q a_i a_j a_q^{-1})
\chi_p (a_{qk}^h a_q a_k a_h^{-1}) = \chi_p(a_{ij}^k a_{qk}^h a_i a_j
a_k a_h^{-1})$ and
$\alpha(\tilde{g_i},\tilde{g_j}\tilde{g_k})\alpha(\tilde{g_j},\tilde{g_k})$
= $\chi_p (a_{jk}^l a_j a_k a_l^{-1}) \chi_p (a_{il}^h a_i a_l a_h^{-1})$
= $\chi_p (a_{il}^h a_{jk}^l a_i a_j a_k a_h^{-1})$
However, because 
$(g_i g_j) g_k =g_q a_{ij}^q g_k= g_h a_{ij}^q a_{qk}^h
=g_i (g_j g_k)= g_i g_l a_{jk}^l = g_h a_{il}^h a_{jk}^l$ 
we have $a_{ij}^k a_{qk}^h=a_{il}^h a_{jk}^l$, and so
\[
(ii)\quad\alpha(\tilde{g_i},\tilde{g_j})\alpha(\tilde{g_i}\tilde{g_j},\tilde{g_k})
=\alpha(\tilde{g_i},\tilde{g_j}\tilde{g_k})\alpha(\tilde{g_j},\tilde{g_k}).
\]
Let us summarize the result. To read out the cocycle according to
\eref{cocycle} we need only two pieces of information: 
the choices of the
representative element in $G^*$ (i.e., $a_i\in A$), and
the definitions of $G^*$ which allows us to calculate the
$a_{ij}^k \in A$. We do not even need to calculate the character
table of $G^*$ to obtain the cocycle. Moreover, in a recent paper \cite{Craps} 
the values of cocycles are being used to construct 
boundary states. We hope our method shall make this above construction
easier.
\section{Conclusions and Prospects}
With the advent of discrete torsion in string theory, the hitherto
novel subject of projective representations has breathed out its
fragrance from mathematics into physics. However a short-coming has been
immediate: the necessary tools for physical computations have so far
been limited in the community due to the unavoidable fact that they,
if present in the mathematical literature, are obfuscated under often
too-technical theorems.

It has been the purpose of this writing, a companion to \cite{FHHP},
to diminish the mystique of projective reprsentations in the context
of constructing gauge theories on D-branes probing orbifolds with
discrete torsion (non-trivial NS-NS B-fields) turned on.
In particular we have deviced an algorithm (Subsection 4.3),
culminating into Theorem 4.4, which computes the gauge theory data 
of the orbifold theory. The advantage of the method is its
directness: without recourse to the sophistry of twisted group algebras
and projective characters as had been suggested by some recent works
\cite{Doug,AspinPles}, all methods so-far known in the treatment of
orbifolds (e.g. \cite{LNV,HanHe}) are immediately generalisable.

We have shown that in computing the matter spectrum for an orbifold
$G$ with discrete torsion turned on, all that is
required is the ordinary charater table char$(G^*)$ of the covering
group $G^*$ of $G$. This table, together with the available character
table of $G$, immediately gives a quiver diagram which splits into
$|M(G)|$ disjoint pieces ($M(G)$ is the Schur Multiplier of
$G$), one of which is the ordinary quiver for $G$ and the rest, are
precisely the quivers for the various non-trivial discrete torsions.

A host of examples are then presented, demonstrating the systematic
power of the algorithm. In particular we have tabulated the results
for all the exceptional subgroups of $SU(3)$ as well as some first
members of the $\Delta$-series.

Directions for future research are self-evident. Brane setups for
orbifolds with discrete torsion have yet to be established. We
therefore need to investigate the groups satisfying BBM condition as
defined in \cite{ZD}, such as the intransitives of the form $\IZ
\times \IZ$ and $\IZ \times D$. Furthermore, we have given the
presentation of the covering groups of series such as $\IZ \times
\IZ$, $\IZ\times D,
\IZ\times E$ and $\Delta(3n^2), \Delta(6n^2)$. It will be interesting
to find the analytic results of the possible quivers.

More important, as we have reduced the problem of orbifolds with discrete
torsion to that of {\em linear} representations, we can instantly
extend the methods of \cite{LNV} to compute superpotentials and thence
further to an extensive and systematic study of non-commutative moduli
spaces in the spirit of \cite{BJL}.
So too do the families of toric varieties await us, methods utilised
in \cite{Tatar,Toric} eagerly anticipate their extension.
%Indeed we have set a vessel adrift, it shall take the course in a vast
%and unknown sea.
%
\section*{Acknowledgements}
%We would especially like to dis-acknowledge the computer
%administration of LNS for the dis-service and inconveniences they have
%caused in the preparation of this manuscript.
{\it Ad Catharinae Sanctae Alexandriae et Ad Majorem Dei Gloriam...\\}
We would like to extend our sincere gratitude to Professor J. Humphreys of
the University of Liverpool, UK for his helpful insight in projective
characters. Furthermore, we are very much obliged to D. Berenstein for
helpful discussion and his careful examinations and corrections to
earlier versions of the manuscript.
Finally we gratefully acknowledge the CTP of MIT as
well as Dr.~Charles Reed for their gracious patronage.
\newpage
\section{Appendix}
We here present, for the reference of the reader, the (ordinary)
character tables of the groups as well as the covering groups thereof,
of the examples which we studied in Section 4.
\vspace{2.0cm}
\[
{\tiny
\Sigma(60) \quad
\ba{|c|c|c|c|c|}
\hline
1 & 12 & 12 & 15 & 20 \\ \hline 1 & 1 & 1 & 1 & 1 \\ \hline 3 & -
\omega_5^2 - \omega_5^{-2} & -\omega_5 - \omega_5^{-1} & -1 & 0 \\ \hline 3 & -\omega_5 - 
    \omega_5^{-1} & -\omega_5^2 - \omega_5^{-2} & -1 & 0 \\ \hline 4 & 
     -1 & -1 & 0 & 1 \\ \hline 5 & 0 & 0 & 1 & -1 \\ \hline
\ea
\qquad
\Sigma(60)^* \quad
\ba{|c|c|c|c|c|c|c|c|c|}
\hline
    1 & 1 & 12 & 12 & 12 & 12 & 30 & 20 & 20 \\ \hline 1 & 1 & 1 & 1 & 1 & 1 & 1 & 
    1 & 1 \\ \hline 3 & 3 & -\omega_5^2 - \omega_5^{-2} & -\omega_5^2 - 
    \omega_5^{-2} & -\omega_5 - \omega_5^{-1} & -\omega_5 - \omega_5^{-1} & 
     -1 & 0 & 0 \\ \hline 3 & 3 & -\omega_5 - \omega_5^{-1} & -\omega_5 - \omega_5^{-1} & -
      \omega_5^2 - \omega_5^{-2} & -\omega_5^2 - 
    \omega_5^{-2} & -1 & 0 & 0 \\ \hline 4 & 4 & -1 & -1 & -1 & 
     -1 & 0 & 1 & 1 \\ \hline 5 & 5 & 0 & 0 & 0 & 0 & 1 & -1 & -1 \\ \hline 2 & -2 & 
      -\omega_5^2 - \omega_5^{-2} & \omega_5^2 + 
    \omega_5^{-2} & -\omega_5 - \omega_5^{-1} & \omega_5 + \omega_5^{-1} & 0 & 
     -1 & 1 \\ \hline 2 & -2 & -\omega_5 - \omega_5^{-1} & \omega_5 + \omega_5^{-1} & -
      \omega_5^2 - \omega_5^{-2} & \omega_5^2 + 
    \omega_5^{-2} & 0 & -1 & 1 \\ \hline 4 & -4 & 1 & -1 & 1 & -1 & 0 & 1 & 
     -1 \\ \hline 6 & -6 & -1 & 1 & -1 & 1 & 0 & 0 & 0 \\ \hline
\ea
}
\]

\[
{\tiny
\Sigma(168) \quad
\ba{|c|c|c|c|c|c|}
\hline
1 & 21 & 42 & 56 & 24 & 24 \\ \hline 1 & 1 & 1 & 1 & 1 & 1 \\ \hline 3 & 
     -1 & 1 & 0 & a & \bar{a} \\ \hline 3 & -1 & 1 & 0 & \bar{a}
     & a \\ \hline 6 & 2 & 0 & 0 & -1 & -1 \\ \hline 7 & -1 & 
     -1 & 1 & 0 & 0 \\ \hline 8 & 0 & 0 & -1 & 1 & 1 \\ \hline
\ea
\qquad
\Sigma(168)^* \quad
\ba{|c|c|c|c|c|c|c|c|c|c|c|}
\hline
    1 & 1 & 42 & 42 & 42 & 56 & 56 & 24 & 24 & 24 & 24 \\ \hline 1 & 
    1 & 1 & 1 & 1 & 1 & 1 & 1 & 1 & 1 & 1 \\ \hline 3 & 3 & 
     -1 & 1 & 1 & 0 & 0 & a & a & \bar{a} & \bar{a}
     \\ \hline 3 & 3 & -1 & 1 & 1 & 0 & 0 & \bar{a} & \bar{a}
     & a & a \\ \hline 6 & 6 & 2 & 0 & 0 & 0 & 0 & -1 & -1 & -1 & 
     -1 \\ \hline 7 & 7 & -1 & -1 & 
     -1 & 1 & 1 & 0 & 0 & 0 & 0 \\ \hline 8 & 8 & 0 & 0 & 0 & -1 & 
     -1 & 1 & 1 & 1 & 1 \\ \hline 4 & -4 & 0 & 0 & 0 & 1 & -1 & 
     -a & a & -\bar{a} & \bar{a} \\ \hline 4 & 
     -4 & 0 & 0 & 0 & 1 & -1 & -\bar{a} & \bar{a} & 
     -a & a \\ \hline 6 & -6 & 0 & -{\sqrt{2}} & {\sqrt{2}} & 0 & 0 & 
     -1 & 1 & -1 & 1 \\ \hline 6 & -6 & 0 & {\sqrt{2}} & -{
       \sqrt{2}} & 0 & 0 & -1 & 1 & -1 & 1 \\ \hline 8 & -8 & 0 & 0 & 0 & 
     -1 & 1 & 1 & -1 & 1 & -1 \\ \hline
\ea
\qquad
a := \frac{-1 + \sqrt{7} i}{2}
}
\]

\[
{\tiny
\Sigma(1080) \quad
\ba{|c|c|c|c|c|c|c|c|c|c|c|c|c|c|c|c|c|}
\hline
1 & 1 & 1 & 45 & 45 & 45 & 72 & 72 & 72 & 72 & 72 & 72 & 90 & 90 & 90 & 120 & 120 \\ \hline 1 & 1 & 1 & 1 & 1 & 1 & 1 & 1 & 1 & 1 & 1 & 1 & 1 & 1 & 1 & 1 & 1 \\ \hline 3 &
   3\bar{A} & 3A & -A & -\bar{A} & -1 & X & Y & Z & W & \bar{Z} & \bar{W} & \bar{A} & A & 1 & 0 & 0 \\ \hline 3 & 3
   \bar{A} & 3A & -A & -\bar{A} & -1 & Y & X & W & Z & \bar{W} & \bar{Z} & \bar{A} & A & 1 & 0 & 0 \\ \hline 3 & 3
   A & 3\bar{A} & -\bar{A} & -A & -1 & X & Y & \bar{Z} & \bar{W} & Z & W & A & \bar{A} & 1 & 0 & 0 \\ \hline 3 & 3
   A & 3\bar{A} & -\bar{A} & -A & -1 & Y & X & \bar{W} & \bar{Z} & W & Z & A & \bar{A}
    & 1 & 0 & 0 \\ \hline 5 & 5 & 5 & 1 & 1 & 1 & 0 & 0 & 0 & 0 & 0 & 0 & -1 & -1 & -1 & 2 & -1 \\ \hline 5 & 5 & 5 & 1 & 1 & 1 & 0 & 0 & 0 & 0 & 0 & 0 & -1 & -1 & -1 & 
    -1 & 2 \\ \hline 6 & 6\bar{A} & 6A & 2A & 2\bar{A} & 2 & 1 & 1 & \bar{A} & \bar{A}
    & A & A & 0 & 0 & 0 & 0 & 0 \\ \hline 6 & 6A & 6\bar{A} & 2\bar{A} & 2A & 2 & 1 & 1 & A & A & \bar{A} & \bar{A} & 0 & 0 & 0 & 0 & 0 \\ \hline 8 & 8 & 8 & 0 & 0 & 0 & X & Y & Y & X & Y & X & 0 & 0 & 0 & -1 & -1 \\ \hline 8 & 8 & 8 & 0 & 0 & 0 & Y & X & X & Y & X & Y & 0 & 0 & 0 & 
    -1 & -1 \\ \hline 9 & 9 & 9 & 1 & 1 & 1 & -1 & -1 & -1 & -1 & -1 & -1 & 1 & 1 & 1 & 0 & 0 \\ \hline 9 & 9\bar{A} & 9A & A & \bar{A} & 1 & -1 & 
    -1 & -\bar{A} & -\bar{A} & -A & -A & \bar{A} & A & 1 & 0 & 0 \\ \hline 9 & 9A & 9\bar{A} & \bar{A} & A & 1 & -1 & 
    -1 & -A & -A & -\bar{A} & -\bar{A} & A & \bar{A} & 1 & 0 & 0 \\ \hline 10 & 10 & 10 & -2 & -2 & 
    -2 & 0 & 0 & 0 & 0 & 0 & 0 & 0 & 0 & 0 & 1 & 1 \\ \hline 15 & 15\bar{A} & 15A & -A & -\bar{A} & -1 & 0 & 0 & 0 & 0 & 0 & 0 & -\bar{A} & -A & -1 & 0 & 0 \\ \hline 15 & 15A & 15\bar{A} & -\bar{A} & -A & -1 & 0 & 0 & 0 & 0 & 0 & 0 & -A & -\bar{A} & -1 & 0 & 0 \\ \hline
\ea
\qquad
\ba{l}
A := \omega_3;\\
B := \omega_5;\\
C := \omega_{15};\\
X := -B-\bar{B};\\
Y := -B^2 - \bar{B}^2;\\
Z := -C - C^4;\\
W := -\bar{C}^2 - C^7;
\ea
}
\]

\newpage
$\Sigma(1080)^*$
{\tiny 
\[
\vspace{-0.2in}
D := B+\bar{B}, E := B^2 + \bar{B}^2, F
:= \bar{C} + \bar{C}^4, G := C^2 + \bar{C}^7, H := \omega_{24}, J :=
\bar{H}^7 - H^{11}, K := \bar{H}^5 - H
\]
}
\[
{\tiny
\ba{|c|c|c|c|c|c|c|c|c|c|c|c|c|c|c|c|c|c|c|c|c|c|c|c|c|c|c|c|c|c|c|}
\hline
   1 & 1 & 1 & 1 & 1 & 1 & 90 & 90 & 90 & 72 & 72 & 72 & 72 & 72 & 72 & 72 & 72 & 72 & 72 & 72 & 72 & 90 & 90 & 90 & 90 & 90 & 90 & 120 & 120 & 120 & 120 \\ \hline 1 & 
   1 & 1 & 1 & 1 & 1 & 1 & 1 & 1 & 1 & 1 & 1 & 1 & 1 & 1 & 1 & 1 & 1 & 1 & 1 & 1 & 1 & 1 & 1 & 1 & 1 & 1 & 1 & 1 & 1 & 1 \\ \hline 3 & 3 & 3\bar{A} & 3
   \bar{A} & 3A & 3A & -A & -\bar{A} & -1 & X & X & Y & Y & Z & Z & W & W & \bar{Z} & \bar{Z} & \bar{W} & 
   \bar{W} & \bar{A} & \bar{A} & A & A & 1 & 1 & 0 & 0 & 0 & 0 \\ \hline 3 & 3 & 3\bar{A} & 3\bar{A} & 3A & 3A & 
    -A & -\bar{A} & -1 & Y & Y & X & X & W & W & Z & Z & \bar{W} & \bar{W} & \bar{Z} & \bar{Z} & \bar{A} & \bar{A} & A & A & 1 & 1 & 0 & 0 & 0 & 0 \\ \hline 3 & 3 & 3A & 3A & 3\bar{A} & 3\bar{A} & -\bar{A} & -A & 
    -1 & X & X & Y & Y & \bar{Z} & \bar{Z} & \bar{W} & \bar{W} & Z & Z & W & W & A & A & \bar{A} & \bar{A} & 1 & 1 & 0 & 0 & 0 & 0 \\ \hline 3 & 3 & 3A & 3A & 3\bar{A} & 3\bar{A} & -\bar{A} & -A & -1 & Y & Y & X & X & \bar{W} & \bar{W} & \bar{Z} & \bar{Z} & W & W & Z & Z & A & A & \bar{A} & \bar{A} & 1 & 1 & 0 & 0 & 0 & 0 \\ \hline 5 & 5 & 5 & 5 & 5 & 5 & 1 & 1 & 1 & 0 & 0 & 0 & 0 & 0 & 0 & 0 & 0 & 0 & 0 & 0 & 0 & -1 & -1 & -1 & -1 & -1 & -1 & 2 & 2 & -1 & 
    -1 \\ \hline 5 & 5 & 5 & 5 & 5 & 5 & 1 & 1 & 1 & 0 & 0 & 0 & 0 & 0 & 0 & 0 & 0 & 0 & 0 & 0 & 0 & -1 & -1 & -1 & -1 & -1 & -1 & -1 & -1 & 2 & 2 \\ \hline 6 & 6 & 6
   \bar{A} & 6\bar{A} & 6A & 6A & 2A & 2\bar{A} & 2 & 1 & 1 & 1 & 1 & \bar{A} & \bar{A} & \bar{A} & \bar{A} & A & A & A & A & 0 & 0 & 0 & 0 & 0 & 0 & 0 & 0 & 0 & 0 \\ \hline 6 & 6 & 6A & 6A & 6\bar{A} & 6\bar{A} & 2
   \bar{A} & 2A & 2 & 1 & 1 & 1 & 1 & A & A & A & A & \bar{A} & \bar{A} & \bar{A} & \bar{A} & 0 & 0 & 0 & 0 & 0 & 0 & 0 & 0 & 0 & 0 \\ \hline 8 & 8 & 8 & 8 & 8 & 8 & 0 & 0 & 0 & X & X & Y & Y & Y & Y & X & X & Y & Y & X & X & 0 & 0 & 0 & 0 & 0 & 0 & -1 & 
    -1 & -1 & -1 \\ \hline 8 & 8 & 8 & 8 & 8 & 8 & 0 & 0 & 0 & Y & Y & X & X & X & X & Y & Y & X & X & Y & Y & 0 & 0 & 0 & 0 & 0 & 0 & -1 & -1 & -1 & 
    -1 \\ \hline 9 & 9 & 9 & 9 & 9 & 9 & 1 & 1 & 1 & -1 & -1 & -1 & -1 & -1 & -1 & -1 & -1 & -1 & -1 & -1 & -1 & 1 & 1 & 1 & 1 & 1 & 1 & 0 & 0 & 0 & 0 \\ \hline 9 & 9 & 9
   \bar{A} & 9\bar{A} & 9A & 9A & A & \bar{A} & 1 & -1 & -1 & -1 & -1 & -\bar{A} & -\bar{A} & -\bar{A} & -\bar{A} & -A & -A & -A & -A & \bar{A} & \bar{A} & A & A & 1 & 1 & 0 & 0 & 0 & 0 \\ \hline 9 & 9 & 9A & 9A & 9
   \bar{A} & 9\bar{A} & \bar{A} & A & 1 & -1 & -1 & -1 & -1 & -A & -A & -A & -A & -\bar{A} & -\bar{A} & -\bar{A} & -\bar{A} & A & A & \bar{A} & \bar{A} & 1 & 1 & 0 & 0 & 0 & 0 \\ \hline 10 & 10 & 10 & 10 & 10 & 10 & -2 & -2 & 
    -2 & 0 & 0 & 0 & 0 & 0 & 0 & 0 & 0 & 0 & 0 & 0 & 0 & 0 & 0 & 0 & 0 & 0 & 0 & 1 & 1 & 1 & 1 \\ \hline 15 & 15 & 15\bar{A} & 15\bar{A} & 15
   A & 15A & -A & -\bar{A} & -1 & 0 & 0 & 0 & 0 & 0 & 0 & 0 & 0 & 0 & 0 & 0 & 0 & -\bar{A} & -\bar{A} & -A & -A & -1 & 
    -1 & 0 & 0 & 0 & 0 \\ \hline 15 & 15 & 15A & 15A & 15\bar{A} & 15\bar{A} & -\bar{A} & -A & 
    -1 & 0 & 0 & 0 & 0 & 0 & 0 & 0 & 0 & 0 & 0 & 0 & 0 & -A & -A & -\bar{A} & -\bar{A} & -1 & -1 & 0 & 0 & 0 & 0 \\ \hline 4 & -4 & -4 & 4 & 
    -4 & 4 & 0 & 0 & 0 & 1 & -1 & 1 & -1 & 1 & -1 & 1 & -1 & 1 & -1 & 1 & -1 & 0 & 0 & 0 & 0 & 0 & 0 & 1 & -1 & -2 & 2 \\ \hline 4 & -4 & -4 & 4 & 
    -4 & 4 & 0 & 0 & 0 & 1 & -1 & 1 & -1 & 1 & -1 & 1 & -1 & 1 & -1 & 1 & -1 & 0 & 0 & 0 & 0 & 0 & 0 & -2 & 2 & 1 & -1 \\ \hline 6 & -6 & -6A & 6A & -6
   \bar{A} & 6\bar{A} & 0 & 0 & 0 & -1 & 1 & -1 & 1 & -A & A & -A & A & -\bar{A} & \bar{A} & -\bar{A} & \bar{A} & J & -J & K & -K & -{\sqrt{2}} & {\sqrt{2}} & 0 & 0 & 0 & 0 \\ \hline 6 & -6 & -6A & 6A & -6\bar{A} & 6\bar{A} & 0 & 0 & 0 & -1 & 1 & 
    -1 & 1 & -A & A & -A & A & -\bar{A} & \bar{A} & -\bar{A} & \bar{A} & -J & J & -K & K & {\sqrt{2}} & -{
      \sqrt{2}} & 0 & 0 & 0 & 0 \\ \hline 6 & -6 & -6\bar{A} & 6\bar{A} & -6A & 6A & 0 & 0 & 0 & -1 & 1 & -1 & 1 & -\bar{A} & 
   \bar{A} & -\bar{A} & \bar{A} & -A & A & -A & A & -K & K & -J & J & -{\sqrt{2}} & {\sqrt{2}} & 0 & 0 & 0 & 0 \\ \hline 6 & -6 & -6
   \bar{A} & 6\bar{A} & -6A & 6A & 0 & 0 & 0 & -1 & 1 & -1 & 1 & -\bar{A} & \bar{A} & -\bar{A} & \bar{A} & -A & A & -A & A & K & -K & J & -J & {\sqrt{2}} & -{\sqrt{2}} & 0 & 0 & 0 & 0 \\ \hline 8 & -8 & -8 & 8 & 
    -8 & 8 & 0 & 0 & 0 & D & X & E & Y & E & Y & D & X & E & Y & D & X & 0 & 0 & 0 & 0 & 0 & 0 & -1 & 1 & -1 & 1 \\ \hline 8 & -8 & -8 & 8 & 
    -8 & 8 & 0 & 0 & 0 & E & Y & D & X & D & X & E & Y & D & X & E & Y & 0 & 0 & 0 & 0 & 0 & 0 & -1 & 1 & -1 & 1 \\ \hline 10 & -10 & -10 & 10 & 
    -10 & 10 & 0 & 0 & 0 & 0 & 0 & 0 & 0 & 0 & 0 & 0 & 0 & 0 & 0 & 0 & 0 & -{\sqrt{2}} & {\sqrt{2}} & {\sqrt{2}} & -{\sqrt{2}} & -{\sqrt{2}} & {\sqrt{2}} & 1 & 
    -1 & 1 & -1 \\ \hline 10 & -10 & -10 & 10 & -10 & 10 & 0 & 0 & 0 & 0 & 0 & 0 & 0 & 0 & 0 & 0 & 0 & 0 & 0 & 0 & 0 & {\sqrt{2}} & -{\sqrt{2}} & -{\sqrt{2}} & {
     \sqrt{2}} & {\sqrt{2}} & -{\sqrt{2}} & 1 & -1 & 1 & -1 \\ \hline 12 & -12 & -12A & 12A & -12\bar{A} & 12
   \bar{A} & 0 & 0 & 0 & X & D & Y & E & \bar{Z} & \bar{F} & \bar{W} & G & Z & F & W & \bar{G} & 0 & 0 & 0 & 0 & 0 & 0 & 0 & 0 & 0 & 0 \\ \hline 12 & -12 & -12A & 12A & -12\bar{A} & 12\bar{A} & 0 & 0 & 0 & Y & E & X & D & \bar{W} & G & \bar{Z} & \bar{F} & W & \bar{G} & Z & F & 0 & 0 & 0 & 0 & 0 & 0 & 0 & 0 & 0 & 0 \\ \hline 12 & -12 & -12
   \bar{A} & 12\bar{A} & -12A & 12A & 0 & 0 & 0 & X & D & Y & E & Z & F & W & \bar{G} & \bar{Z} & \bar{F} & 
   \bar{W} & G & 0 & 0 & 0 & 0 & 0 & 0 & 0 & 0 & 0 & 0 \\ \hline 12 & -12 & -12\bar{A} & 12\bar{A} & -12A & 12
   A & 0 & 0 & 0 & Y & E & X & D & W & \bar{G} & Z & F & \bar{W} & G & \bar{Z} & \bar{F} & 0 & 0 & 0 & 0 & 0 & 0 & 0 & 0 & 0 & 0 \\ \hline
\ea
}
\]

\[
{\tiny
\Delta(6 \times 2^2) = \quad
\ba{|c|c|c|c|c|}
\hline
1 & 3 & 6 & 6 & 8 \\ \hline 1 & 1 & 1 & 1 & 1 \\ \hline 1 & 1 & 
    -1 & -1 & 1 \\ \hline 2 & 2 & 0 & 0 & -1 \\ \hline 3 & -1 & 
    -1 & 1 & 0 \\ \hline 3 & -1 & 1 & -1 & 0 \\ \hline 
\ea
\qquad
\Delta(6\times 2^2)^* = \quad
\ba{|c|c|c|c|c|c|c|c|}
\hline
1 & 1 & 6 & 6 & 6 & 12 & 8 & 8 \\ \hline 1 & 1 & 1 & 1 & 1 & 
   1 & 1 & 1 \\ \hline 1 & 1 & 1 & -1 & -1 & 
    -1 & 1 & 1 \\ \hline 2 & 2 & 2 & 0 & 0 & 0 & -1 & 
    -1 \\ \hline 3 & 3 & -1 & -1 & -1 & 1 & 0 & 0 \\ \hline 3 & 3 & 
    -1 & 1 & 1 & -1 & 0 & 0 \\ \hline 2 & -2 & 0 & -e^
     {\frac{i }{4}\pi } - e^{\frac{3i }{4}\pi }
   & e^{\frac{i }{4}\pi } + 
   e^{\frac{3i }{4}\pi } & 0 & -1 & 1 \\ \hline 2 & 
    -2 & 0 & e^{\frac{i }{4}\pi } + 
   e^{\frac{3i }{4}\pi } & -e^{\frac{i }{4}\pi } - e^{\frac{3i }{4}\pi }
   & 0 & -1 & 1 \\ \hline 4 & -4 & 0 & 0 & 0 & 0 & 1 & -1 \\ \hline
\ea
}
\]

\[
\ba{lcl}
\Delta(6 \times 4^2) = & & \Delta(6\times 4^2)^* = \\
{\tiny
\ba{|c|c|c|c|c|c|c|c|c|c|}
\hline
   1 & 3 & 3 & 3 & 6 & 12 & 12 & 12 & 12 & 32 \\ \hline 1 & 1 & 1 & 1 & 1 & 1 & 
    1 & 1 & 1 & 1 \\ \hline 1 & 1 & 1 & 1 & 1 & -1 & -1 & -1 & 
     -1 & 1 \\ \hline 2 & 2 & 2 & 2 & 2 & 0 & 0 & 0 & 0 & -1 \\ \hline 3 & 3 & -1 & 
     -1 & -1 & -1 & 1 & 1 & -1 & 0 \\ \hline 3 & 3 & -1 & -1 & -1 & 1 & -1 & 
     -1 & 1 & 0 \\ \hline 3 & -1 & -1 - 2i  & -1 + 2i  & 1 & -1 & i
     & -i  & 1 & 0 \\ \hline 3 & -1 & -1 + 2i  & -1 - 2i  & 1 & 
     -1 & -i  & i  & 1 & 0 \\ \hline 3 & -1 & -1 - 2i  & -1 + 
    2i  & 1 & 1 & -i  & i  & -1 & 0 \\ \hline 3 & -1 & -1 + 
    2i  & -1 - 2i  & 1 & 1 & i  & -i  & -1 & 0 \\ \hline 6 & 
     -2 & 2 & 2 & -2 & 0 & 0 & 0 & 0 & 0 \\ \hline
\ea
}
&&
{\tiny
\ba{|c|c|c|c|c|c|c|c|c|c|c|c|c|c|c|}
\hline
    1 & 1 & 3 & 3 & 6 & 6 & 12 & 24 & 12 & 12 & 12 & 12 & 24 & 32 & 32 \\ \hline 
    1 & 1 & 1 & 1 & 1 & 1 & 1 & 1 & 1 & 1 & 1 & 1 & 1 & 1 & 1 \\ \hline 1 & 1 & 1 &
    1 & 1 & 1 & 1 & -1 & -1 & -1 & -1 & -1 & 
     -1 & 1 & 1 \\ \hline 2 & 2 & 2 & 2 & 2 & 2 & 2 & 0 & 0 & 0 & 0 & 0 & 0 & 
     -1 & -1 \\ \hline 3 & 3 & 3 & 3 & -1 & -1 & -1 & -1 & 1 & 1 & 1 & 1 & 
     -1 & 0 & 0 \\ \hline 3 & 3 & 3 & 3 & -1 & -1 & -1 & 1 & -1 & -1 & -1 & 
     -1 & 1 & 0 & 0 \\ \hline 3 & 3 & -1 & -1 & -1 - 2i  & -1 + 
    2i  & 1 & -1 & i  & i  & -i  & -i
      & 1 & 0 & 0 \\ \hline 3 & 3 & -1 & -1 & -1 + 2i  & -1 - 
    2i  & 1 & -1 & -i  & -i  & i  & i
     & 1 & 0 & 0 \\ \hline 3 & 3 & -1 & -1 & -1 - 2i  & -1 + 
    2i  & 1 & 1 & -i  & -i  & i  & i  & 
     -1 & 0 & 0 \\ \hline 3 & 3 & -1 & -1 & -1 + 2i  & -1 - 
    2i  & 1 & 1 & i  & i  & -i  & -i  & 
     -1 & 0 & 0 \\ \hline 6 & 6 & -2 & -2 & 2 & 2 & 
     -2 & 0 & 0 & 0 & 0 & 0 & 0 & 0 & 0 \\ \hline 2 & -2 & 
     -2 & 2 & 0 & 0 & 0 & 0 & i {\sqrt{2}} & -i {\sqrt{2}} & 
    -i {\sqrt{2}} & i {\sqrt{2}} & 0 & -1 & 1 \\ \hline 2 & -2 & 
     -2 & 2 & 0 & 0 & 0 & 0 & -i {\sqrt{2}} & i {\sqrt{2}} & 
    i {\sqrt{2}} & -i {\sqrt{2}} & 0 & -1 & 1 \\ \hline 4 & -4 & 
     -4 & 4 & 0 & 0 & 0 & 0 & 0 & 0 & 0 & 0 & 0 & 1 & -1 \\ \hline 6 & -6 & 2 & 
     -2 & 0 & 0 & 0 & 0 & -{\sqrt{2}} & {\sqrt{2}} & -{\sqrt{2}} & {
      \sqrt{2}} & 0 & 0 & 0 \\ \hline 6 & -6 & 2 & -2 & 0 & 0 & 0 & 0 & {
      \sqrt{2}} & -{\sqrt{2}} & {\sqrt{2}} & -{\sqrt{2}} & 0 & 0 & 0 \\ \hline 
\ea
}
\ea
\]

\[
\ba{lcl}
\Delta(3 \times 4^2) = & & \Delta(3 \times 4^2)^* = \\
{\tiny
\ba{|c|c|c|c|c|c|c|c|}
\hline
1 & 3 & 3 & 3 & 3 & 3 & 16 & 16 \\ \hline 1 & 1 & 1 & 1 & 1 & 1 & 1 & 1 \\ \hline 1 &
    1 & 1 & 1 & 1 & 1 & \omega_3 & \bar{\omega_3} \\ \hline 1 & 1 & 1 & 1 & 1 & 1 & 
    \bar{\omega_3} & \omega_3 \\ \hline 3 & -1 & -1 & 3 & -1 & 
     -1 & 0 & 0 \\ \hline 3 & 1 & 1 & -1 & -1 - 2i  & -1 + 
    2i  & 0 & 0 \\ \hline 3 & 1 & 1 & -1 & -1 + 2i  & -1 - 
    2i  & 0 & 0 \\ \hline 3 & -1 - 2i  & -1 + 2i  & 
     -1 & 1 & 1 & 0 & 0 \\ \hline 3 & -1 + 2i  & -1 - 2i  & 
     -1 & 1 & 1 & 0 & 0 \\ \hline
\ea}
&&
{\tiny
\ba{|c|c|c|c|c|c|c|c|c|c|c|c|c|c|c|c|c|c|}
\hline
    1 & 1 & 1 & 1 & 12 & 12 & 6 & 6 & 12 & 12 & 16 & 16 & 16 & 16 & 16 & 16 &
    16 & 16 \\ \hline 1 & 1 & 1 & 1 & 1 & 1 & 1 & 1 & 1 & 1 & 1 & 1 & 1 & 1 & 1 & 
    1 & 1 & 1 \\ \hline 1 & 1 & 1 & 1 & 1 & 1 & 1 & 1 & 1 & 1 & \omega_3 & \omega_3 & \omega_3 & \omega_3 &
    \bar{\omega}_3 & \bar{\omega}_3 & \bar{\omega}_3 & \bar{\omega}_3 \\ \hline 
1 & 1 & 1 & 1 & 1 & 1 & 1 & 1 & 1 & 1 & \bar{\omega}_3 & 
    \bar{\omega}_3 & \bar{\omega}_3 & \bar{\omega}_3 & \omega_3 &
    \omega_3 & \omega_3 & \omega_3 
\\ \hline 3 & 3 & 3 & 3 & -1 & -1 & 3 & 3 & -1 & 
     -1 & 0 & 0 & 0 & 0 & 0 & 0 & 0 & 0 \\ \hline 3 & 3 & 3 & 3 & 1 & 1 & -1 & 
     -1 & -1 - 2i  & -1 + 
    2i  & 0 & 0 & 0 & 0 & 0 & 0 & 0 & 0 \\ \hline 3 & 3 & 3 & 3 & 1 & 1 & 
     -1 & -1 & -1 + 2i  & -1 - 
    2i  & 0 & 0 & 0 & 0 & 0 & 0 & 0 & 0 \\ \hline 3 & 3 & 3 & 3 & -1 - 
    2i  & -1 + 2i  & -1 & 
     -1 & 1 & 1 & 0 & 0 & 0 & 0 & 0 & 0 & 0 & 0 \\ \hline 3 & 3 & 3 & 3 & -1 + 
    2i  & -1 - 2i  & -1 & 
     -1 & 1 & 1 & 0 & 0 & 0 & 0 & 0 & 0 & 0 & 0 \\ \hline 2 & -2 & 2 & 
     -2 & 0 & 0 & 2 & -2 & 0 & 0 & -\omega_3 & \omega_3 & \omega_3 & -\omega_3 & -\bar{\omega}_3 & 
     -\bar{\omega}_3 & \bar{\omega}_3 & \bar{\omega}_3 \\ \hline 2 & 
     -2 & 2 & -2 & 0 & 0 & 2 & -2 & 0 & 0 & -\bar{\omega}_3 & 
\bar{\omega}_3 & \bar{\omega}_3 & -\bar{\omega}_3 & -\omega_3 & -\omega_3 & \omega_3 & \omega_3 \\ \hline 2 & 
     -2 & 2 & -2 & 0 & 0 & 2 & -2 & 0 & 0 & -1 & 1 & 1 & -1 & -1 & 
     -1 & 1 & 1 \\ \hline 6 & -6 & 6 & -6 & 0 & 0 & 
     -2 & 2 & 0 & 0 & 0 & 0 & 0 & 0 & 0 & 0 & 0 & 0 \\ \hline 4 & 4i  & 
     -4 & -4i  & 0 & 0 & 0 & 0 & 0 & 0 & \bar{\omega}_3 & 
-\bar{\omega}_{12} & \bar{\omega}_{12} & -\bar{\omega}_3 & \omega_3 &
-\omega_3 & \bar{\omega}_{12}^5 & 
-\bar{\omega}_{12}^5 \\ \hline 4 & 4i  & -4 & -4
    i  & 0 & 0 & 0 & 0 & 0 & 0 & \omega_3 & -\bar{\omega}_{12}^5 & 
\bar{\omega}_{12}^5 & -\omega_3 & \bar{\omega}_3 & -\bar{\omega}_3 & 
\bar{\omega}_{12} & -\bar{\omega}_{12} \\ \hline 4 & 4i  & -4 & -4
    i  & 0 & 0 & 0 & 0 & 0 & 0 & 1 & -i  & i  & -1 & 1 & 
     -1 & i  & -i  \\ \hline 4 & -4i  & -4 & 4
    i  & 0 & 0 & 0 & 0 & 0 & 0 & \bar{\omega}_3 & \bar{\omega}_{12} 
& -\bar{\omega}_{12} & -\bar{\omega}_3 & \omega_3 & -\omega_3 & -\bar{\omega}_{12}^5 
& \bar{\omega}_{12}^5 \\ \hline 4 & -4i  & -4 & 4
    i  & 0 & 0 & 0 & 0 & 0 & 0 & \omega_3 & \bar{\omega}_{12}^5 &
-\bar{\omega}_{12}^5 & 
-\omega_3 & \bar{\omega}_3 & -\bar{\omega}_3 & -\bar{\omega}_{12} & 
    \bar{\omega}_{12} \\ \hline 4 & -4i  & -4 & 4
    i  & 0 & 0 & 0 & 0 & 0 & 0 & 1 & i  & -i  & -1 & 1 & 
     -1 & -i  & i  \\ \hline 
\ea
}
\ea
\]

\[
\ba{lcl}
\Delta(3 \times 5^2) = && \Delta(3 \times 5^2)^* = \\
{\tiny
\ba{|c|c|c|c|c|c|c|c|c|c|c|}
\hline
    1 & 3 & 3 & 3 & 3 & 3 & 3 & 3 & 3 & 25 & 25 \\ \hline 1 & 1 & 1 & 1 & 1 & 1 & 
    1 & 1 & 1 & 1 & 1 \\ \hline 1 & 1 & 1 & 1 & 1 & 1 & 1 & 1 & 1 & E & \bar{E} \\ \hline 
1 & 1 & 1 & 1 & 1 & 1 & 1 & 1 & 1 & \bar{E} 
& E \\ \hline 3 & A & A & B & B & C & D & \bar{C} & \bar{D} & 0 & 0 \\ \hline 
3 & A & A & B & B & \bar{C} & \bar{D} & C & D & 0 & 0 \\ \hline 
3 & B & B & A & A & \bar{D} & C & D & 
    \bar{C} & 0 & 0 \\ \hline 3 & B & B & A & A & D & \bar{C} & \bar{D} 
& C & 0 & 0 \\ \hline 3 & C & \bar{C} & D & 
    \bar{D} & B & A & B & A & 0 & 0 \\ \hline 3 & \bar{D} & D & C & 
\bar{C} & A & B & A & B & 0 & 0 \\ \hline 3 & \bar{C} & C & \bar{D} 
& D & B & A & B & A & 0 & 0 \\ \hline 3 & D & 
\bar{D} & \bar{C} & C & A & B & A & B & 0 & 0 \\ \hline
\ea}
&&
{\tiny
\ba{|c|c|c|c|c|c|c|c|c|c|c|c|c|c|c|c|c|c|c|c|c|c|c|}
\hline
   1 & 1 & 1 & 1 & 1 & 15 & 15 & 15 & 15 & 15 & 15 & 15 & 15 & 25 & 25
& 25 & 25 & 25 & 25 & 25 & 25 & 25 & 25 \\
 \hline 1 & 1 & 1 & 1 & 1 & 
   1 & 1 & 1 & 1 & 1 & 1 & 1 & 1 & 1 & 1 & 1 & 1 & 1 & 1 & 1 & 1 & 1 &
1 \\ \hline
 1 & 1 & 1 & 1 & 1 & 1 & 1 & 1 & 1 & 1 & 1 & 1 & 1 & E & 
   E & E & E & E & {\bar{E}} & {\bar{E}} & {\bar{E}} & {\bar{E}} & {\bar{E}
   } \\ \hline 1 & 1 & 1 & 1 & 1 & 1 & 1 & 1 & 1 & 1 & 1 & 1 & 1 & {\bar{E}} & {\bar{E}} & {\bar{E}} & {\bar{E}
   } & {\bar{E}} & E & E & E & E & E \\ \hline 3 & 3 & 3 & 3 & 3 & A & A & B & B & C & D & {\bar{C}} & {\bar{D}
   } & 0 & 0 & 0 & 0 & 0 & 0 & 0 & 0 & 0 & 0 \\ \hline 3 & 3 & 3 & 3 & 3 & A & A & B & B & {\bar{C}} & {\bar{D}
   } & C & D & 0 & 0 & 0 & 0 & 0 & 0 & 0 & 0 & 0 & 0 \\ \hline 3 & 3 &
3 & 3 & 3 & B & B & A & A & {\bar{D}} & C & D & {\bar{C}
   } & 0 & 0 & 0 & 0 & 0 & 0 & 0 & 0 & 0 & 0 \\ \hline 3 & 3 & 3 & 3 & 3 & B & B & A & A & D & {\bar{C}} & {\bar{D}
   } & C & 0 & 0 & 0 & 0 & 0 & 0 & 0 & 0 & 0 & 0 \\ \hline 3 & 3 & 3 & 3 & 3 & C & {\bar{C}} & D & {\bar{D}
   } & B & A & B & A & 0 & 0 & 0 & 0 & 0 & 0 & 0 & 0 & 0 & 0 \\ \hline 3 & 3 & 3 & 3 & 3 & {\bar{D}} & D & C & {\bar{C}
   } & A & B & A & B & 0 & 0 & 0 & 0 & 0 & 0 & 0 & 0 & 0 & 0 \\ \hline 3 & 3 & 3 & 3 & 3 & {\bar{C}} & C & {\bar{D}
   } & D & B & A & B & A & 0 & 0 & 0 & 0 & 0 & 0 & 0 & 0 & 0 & 0 \\ \hline 3 & 3 & 3 & 3 & 3 & D & {\bar{D}} & {\bar{C}
   } & C & A & B & A & B & 0 & 0 & 0 & 0 & 0 & 0 & 0 & 0 & 0 & 0 \\ \hline 5 & 5{\bar{F}^2} & 5F & 5{\bar{F}} & 5
   {\bar{F}^2} & 0 & 0 & 0 & 0 & 0 & 0 & 0 & 0 & -1 & -{\bar{F}} & -{\bar{F}^2} & -{\bar{F}^2} & -F & -1 & -F & 
    -{\bar{F}^2} & -{\bar{F}^2} & -{\bar{F}} \\ \hline 5 & 5{\bar{F}^2} & 5F & 5{\bar{F}} & 5
   {\bar{F}^2} & 0 & 0 & 0 & 0 & 0 & 0 & 0 & 0 & -E & -{\bar{G}^7} & -{\bar{G}} & -{\bar{G}^4} & -{G^2
    } & -{\bar{E}} & -{\bar{G}^7} & -G & -{\bar{G}^4} & -{\bar{G}^2} \\ \hline 5 & 5{\bar{F}^2} & 5F & 5
   {\bar{F}} & 5{\bar{F}^2} & 0 & 0 & 0 & 0 & 0 & 0 & 0 & 0 & -{\bar{E}} & -{\bar{G}^2} & -{\bar{G}^4} & 
    -G & -{\bar{G}^7} & -E & -{G^2} & -{\bar{G}^4} & -{\bar{G}} & -{\bar{G}^7} \\ \hline 5 & 5
   {\bar{F}^2} & 5{\bar{F}} & 5F & 5{\bar{F}^2} & 0 & 0 & 0 & 0 & 0 & 0 & 0 & 0 & -1 & -F & -{\bar{F}^2
    } & -{\bar{F}^2} & -{\bar{F}} & -1 & -{\bar{F}} & -{\bar{F}^2} & -{\bar{F}^2} & -F \\ \hline 5 & 5
   {\bar{F}^2} & 5{\bar{F}} & 5F & 5{\bar{F}^2} & 0 & 0 & 0 & 0 & 0 & 0 & 0 & 0 & -E & -{G^2} & -
    {\bar{G}^4} & -{\bar{G}} & -{\bar{G}^7} & -{\bar{E}} & -{\bar{G}^2} & -{\bar{G}^4} & -G & -
    {\bar{G}^7} \\ \hline 5 & 5{\bar{F}^2} & 5{\bar{F}} & 5F & 5{\bar{F}^2} & 0 & 0 & 0 & 0 & 0 & 0 & 0 & 0 & 
    -{\bar{E}} & -{\bar{G}^7} & -G & -{\bar{G}^4} & -{\bar{G}^2} & -E & -{\bar{G}^7} & -{\bar{G}
    } & -{\bar{G}^4} & -{G^2} \\ \hline 5 & 5{\bar{F}} & 5{\bar{F}^2} & 5{\bar{F}^2} & 5
   F & 0 & 0 & 0 & 0 & 0 & 0 & 0 & 0 & -1 & -{\bar{F}^2} & -F & -{\bar{F}} & -{\bar{F}^2} & -1 & -{\bar{F}^2
    } & -{\bar{F}} & -F & -{\bar{F}^2} \\ \hline 5 & 5{\bar{F}} & 5{\bar{F}^2} & 5{\bar{F}^2} & 5
   F & 0 & 0 & 0 & 0 & 0 & 0 & 0 & 0 & -E & -{\bar{G}} & -{G^2} & -{\bar{G}^7} & -{\bar{G}^4} & -{
    \bar{E}} & -G & -{\bar{G}^2} & -{\bar{G}^7} & -{\bar{G}^4} \\ \hline 5 & 5{\bar{F}} & 5{\bar{F}^2} & 5
   {\bar{F}^2} & 5F & 0 & 0 & 0 & 0 & 0 & 0 & 0 & 0 & -{\bar{E}} & -{\bar{G}^4} & -{\bar{G}^7} & -{\bar{G}^2
    } & -G & -E & -{\bar{G}^4} & -{\bar{G}^7} & -{G^2} & -{\bar{G}} \\ \hline 5 & 5F & 5{\bar{F}^2} & 5
   {\bar{F}^2} & 5{\bar{F}} & 0 & 0 & 0 & 0 & 0 & 0 & 0 & 0 & -1 & -{\bar{F}^2} & -{\bar{F}} & -F & -
    {\bar{F}^2} & -1 & -{\bar{F}^2} & -F & -{\bar{F}} & -{\bar{F}^2} \\ \hline 5 & 5F & 5{\bar{F}^2} & 5
   {\bar{F}^2} & 5{\bar{F}} & 0 & 0 & 0 & 0 & 0 & 0 & 0 & 0 & -E & -{\bar{G}^4} & -{\bar{G}^7} & -{
    G^2} & -{\bar{G}} & -{\bar{E}} & -{\bar{G}^4} & -{\bar{G}^7} & -{\bar{G}^2} & -G \\ \hline 5 & 5F & 5
   {\bar{F}^2} & 5{\bar{F}^2} & 5{\bar{F}} & 0 & 0 & 0 & 0 & 0 & 0 & 0
	& 0 & -{\bar{E}} & -G & -{\bar{G}^2}
	 & -{\bar{G}^7} & -{\bar{G}^4} & -E & -{\bar{G}} & -{G^2} & -{\bar{G}^7} & -{\bar{G}^4
    } \\ \hline
\ea
}
\ea
\]

$
A := -\omega_5 - \bar{\omega}_5, B := -\omega_5^2 - \bar{\omega}_5^2,
C := \bar{\omega}_5 - 2\bar{\omega}_5^2, D := 2\omega_5 +
\bar{\omega}_5^2;
E := \omega_3, F := \bar{\omega}_5, G := \omega_{15}.
$

\bibliographystyle{JHEP}

\end{document}